\documentstyle[12pt,aaspp4]{article}
\def\ltsima{$\; \buildrel < \over \sim \;$}
\def\lsim{\lower.5ex\hbox{\ltsima}}
\def\gtsima{$\; \buildrel > \over \sim \;$}
\def\gsim{\lower.5ex\hbox{\gtsima}}
\def\bi#1{\bbox{#1}}
\def\Kappa{K}
\begin{document}
\title{Ray Tracing Simulations of Weak Lensing by Large-Scale Structure}

\author{Bhuvnesh Jain}
\affil{Dept. of Physics, Johns Hopkins University, Baltimore, MD 21218, USA}
\affil{bjain@pha.jhu.edu}

\author{Uro\v s Seljak}
\affil{Center For Astrophysics, Harvard University, Cambridge, MA
02138, USA}
\affil{Max-Planck-Institut f\"ur Astrophysik, Garching 85740, GERMANY}
\affil{uros@mpa-garching.mpg.de}

\author{Simon White}
\affil{Max-Planck-Institut f\"ur Astrophysik, Garching 85740, GERMANY}
\affil{swhite@mpa-garching.mpg.de}

\def\bi#1{\hbox{\boldmath{$#1$}}} 
\begin{abstract}
We investigate weak lensing by large-scale structure using ray tracing
through N-body simulations.  Photon trajectories are followed through 
high resolution simulations of structure formation to make simulated 
maps of shear and convergence on the sky. Tests with varying numerical
parameters are used to calibrate the accuracy of computed lensing 
statistics on angular scales from $\sim$ 1 arcminute to a few degrees. 
Various aspects of the weak lensing approximation are also tested.
We show that the non-scalar component of the shear generated by the 
multiple deflections is small. For fields a few degrees on a side 
the shear power spectrum is almost entirely in the nonlinear 
regime and agrees well with nonlinear analytical predictions. 
Sampling fluctuations in power spectrum estimates are investigated 
by comparing several ray tracing realizations of a given model. For 
survey areas smaller than a degree on a side the main source of 
scatter is nonlinear coupling to modes larger than the survey. We 
develop a method which uses this effect to estimate $\Omega_{\rm m}$ from 
the scatter in power spectrum estimates for subregions of a larger 
survey. We show that the power spectrum can be measured accurately 
on scales corresponding to $1-10\,  h^{-1}$Mpc with
realistic number densities of source galaxies with large intrinsic 
ellipticities. Non-Gaussian features in the one point distribution 
function of the weak lensing convergence (reconstructed from the 
shear) are also sensitive to $\Omega_{\rm m}$. We suggest several 
techniques for estimating $\Omega_{\rm m}$ in the presence of noise 
and compare their statistical power, robustness and simplicity. 
With realistic number densities of source galaxies 
$\Omega_{\rm m}$ can be determined to within 
$0.1-0.2$ from a deep survey of several square degrees.

\end{abstract}

\keywords{cosmology: theory, cosmology: gravitational lensing, 
methods: numerical}

\section{Introduction}

Mapping the large-scale structure (LSS) of the universe is one of the 
major goals of observational cosmology. Traditionally this is performed 
using large surveys of galaxies. Such surveys map the light
distribution, while most of the matter in the universe appears to be 
dark. This paper focuses on mapping the dark matter using gravitational 
lensing, in particular, weak lensing of distant background galaxies 
by large-scale structure. Weak lensing magnifies and shears the images
of these galaxies, inducing an additional ellipticity which,
although not detectable in any individual image, can be measured as a
function of position on the sky by averaging the ellipticities of 
neighboring galaxies. This observable is determined by a line-of-sight
projection of the mass fluctuations and by the spatial geometry of the
universe. When averaged over windows of arcminute scale, the
ellipticities of distant galaxies probe mass fluctuations on scales 
of 1-10 $h^{-1}$ Mpc. If the shape of the mass fluctuation spectrum
is known, such measures are sensitive to the cosmological parameters 
$\Omega_{\rm m}$ and $\Omega_\Lambda$. Thus while strong lensing leads
to highly distorted or multiple images and probes the densest regions 
of the universe, the inner cores of massive halos, weak lensing
provides a direct measure of typical mass fluctuations on larger scales. 

Observatonal attempts at detection of weak lensing by large-scale structure
include the work of Fahlman et al. (1994), Mould et al. (1994), 
Refregier et al. (1998) and Schneider et al. (1998). 
While a definitive detection has yet to be made, a few groups
have demonstrated the capability to measure the small shear signal
induced by large-scale structure. Kaiser et al. (1998) have made
a convincing detection of coherent shear due to a supercluster. 
Several observational efforts to measure weak lensing that use 
wide field CCD detectors are currently underway (see Mellier 1998 for
a review). 

The first weak lensing calculations to use modern models for the 
large-scale matter distribution were those of \cite{Blandford91,Escude91} 
and \cite{Kaiser92}, based on the pioneering work of Gunn 
(1967). More recently, Villumsen (1996), Stebbins (1996), Bernardeau 
et al (1997) and Kaiser (1998) have made linear calculations for a
variety of cosmologies, while Jain \& Seljak (1997) have included the 
effect of the nonlinear evolution of the matter power spectrum. 
Numerical simulations of gravitational lensing have also been
developed in parallel. Different approximations have been used to 
represent the lumpy distribution of dark matter, typically 
in conjunction with methods which trace rays through a series of
lens planes (Schneider \& Weiss 1988;  Jaroszy\'nski et al 1990; 
Lee \& Paczy\'nski  1990; Jaroszy\'nski 1991; Babul \& Lee 1991;
Bartelmann \& Schneider 1991; Blandford et al. 1991). The 
most detailed numerical study of lensing by large-scale
structure has been conducted by Wambsganss, Cen \& Ostriker (1998), 
building on the earlier work of Wambsganss et al (1995, 1997).
Their work focused on somewhat smaller scales than we consider in 
this paper, and as a result was unable to address some of the
nonlinear mode-coupling effects we are concerned with here. Nevertheless,
there is still considerable overlap with the regime we study. Other
recent studies using ray tracing have been conducted by Premadi, 
Martel \& Matzner (1998), van Waerbeke, Bernardeau \& Mellier (1998),
Bartelmann et al (1998) and Couchman, Barber \& Thomas (1998). 

In this paper we apply a newly developed numerical method to investigate a 
variety of statistics that can be extracted from weak lensing data.
Our numerical approach is similar to that of some of the above references, 
in that we approximate the matter distribution between observer and 
source as a sequence of discrete planes 
(see also Bartelmann \& Schneider 1991). The projected distribution
on these planes is obtained from high resolution N-body simulations 
and is converted to deflection angle and shear matrix
distributions using Fourier techniques. These can then be used to
follow $\gsim 10^{6}$ rays from the observer back to the source plane,
giving the magnification, shear and true source position in each
observed direction. The magnification and shear maps are smoothed on 
different scales and used to construct one-point probability
distribution functions and their low-order moments. The two-point 
correlation function provides another simple and robust statistic 
and, for models with Gaussian initial conditions, it contains all the 
fluctuation information on linear scales. 
We show that to reach the linear regime one must survey regions at
least 10 degrees on a side. This is difficult with existing CCD 
cameras which typically have field diameters of $0.25^{\circ}-
0.5^{\circ}$. Most of the information in the initial surveys 
carried out with these instruments will therefore be in the mildly 
to strongly nonlinear regime; in the following we discuss methods for 
power spectrum extraction from such data.

While nonlinear evolution may limit the information that can be extracted 
from a power spectrum analysis, it also leads to non-Gaussian effects 
that can be detected by statistics other than second moments. These 
effects can provide qualitatively new information because the boundary
of the linear regime depends on the relative density fluctuation, 
$\delta({\bi r})=\rho({\bi r})/\bar{\rho}-1$, while the amplitude of
the weak lensing signal is determined by a projection of the total
mass fluctuation, $\bar{\rho}\delta({\bi r})$. Studies of two-point 
correlators cannot separate the effect of $\bar{\rho}$ from a
variation in the normalization of $\delta$, but the transition to
nonlinearity depends on the density contrast alone, so that the
behavior of higher moments allows a separate determination of the density 
parameter $\Omega_{\rm m}=\bar{\rho}/\rho_{\rm crit}$ 
and the matter power spectrum $P_{\delta}(k)$ (\cite{Bernardeau97,JS97}).

\begin{figure}[t!]
\vspace*{6.3cm}
\caption{The probability distribution function (pdf) 
of the convergence $\kappa$ is compared with the pdf of
$\delta$, the 3-dimensional density contrast. 
The right panel shows the pdf of $\delta$ for an open CDM model (dashed)  
and an Einstein-de Sitter CDM model (solid), both smoothed
on a scale of 3 $h^{-1}$ Mpc. The rms density contrast in the two 
models is nearly identical. The left panel shows the pdf of $\kappa$ 
for the same two models. The thick solid curve uses the same smoothing
scale as the dashed curve, 0.5 arcmin, while the thin solid curve 
uses a somewhat larger smoothing scale chosen to give the same rms 
as the dashed curve. The rms approximately 
determines the full pdf for $\delta$ but 
not for $\kappa$. 
}
\includegraphics{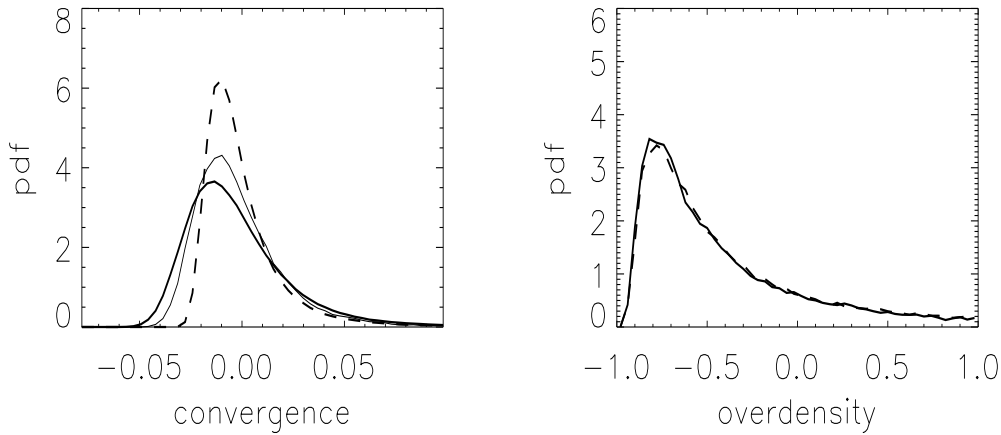}
\label{figdeltapdf}
\end{figure}

In this respect weak lensing maps are easier to interpret than maps of
the galaxy distribution, where non-Gaussian signatures reflect both
nonlinear gravitational clustering and the details of the relation 
between the galaxy distribution and the underlying mass distribution. 
Checks of the consistency of any assumed ``bias'' relation are 
certainly valuable, but they do not provide qualitatively new 
information on cosmological parameters; that is why dynamical measures 
such as streaming motion or velocity dispersion estimates are needed
for galaxy surveys to provide a robust probe of 
$\Omega_{\rm m}$. This point is illustrated in figure \ref{figdeltapdf},
which compares the one point distribution functions 
(pdf) of the convergence $\kappa$ and the density contrast $\delta$
for open and Einstein-de Sitter universes both with CDM power
spectrum shape parameter $\Gamma=0.2$. (These were made from the
ray-tracing simulations discussed in \S 3.) The pdf's of $\kappa$ 
differ substantially even when their rms values are the same. For
$\delta$, on the other hand, the rms completely determines the pdf,
leaving no leverage to determine $\Omega_{\rm m}$. 

In this paper we investigate the observability of 
nonlinear effects in weak lensing data. We apply 
statistics which  measure a variety of non-Gaussian signatures to 
simulated data. These are obtained by ray-tracing through $N$-body 
simulations, and sampling the resulting shear fields with realistic
number densities of source galaxies. Our main goal is to see how 
accurately  the density parameter can be estimated, given the
uncertainty in other parameters, such as the shape and amplitude of the 
power spectrum. The main statistic we investigate is the one-point 
pdf of the convergence or projected matter density. It can be analyzed in many 
different ways, for example by using direct maximum likelihood (ML)
parameter estimation, by measuring various moments, or by
constructing an Edgeworth expansion. In principle ML
should give the most powerful constraints, but it is somewhat 
cumbersome to use and it requires a suite of cosmological simulations 
spanning the parameter space considered. The latter two
methods are simpler and can be tested against perturbation theory, 
but they use the information in a less optimal way. In the following, 
we evaluate the performance of these estimators and compare their 
robustness, statistical power and simplicity.

Our results on the second moment are valid for both the convergence
and shear, as their second moments are identical in the weak lensing
approximation and for the range of scales of interest for lensing by
large-scale structure. For the skewness and the pdf, we present results
for the convergence, which is not a direct observable. We demonstrate
that for fields larger than a degree on a side it is possible to reconstruct
the convergence from data on the ellipticities of galaxies by using the
Fourier space relations between convergence and shear 
(equation \ref{kappaft}). In our modeling
of observational data we have included the random errors in shear
measurement due to the finite number of source galaxies, each of which
has a large intrinsic ellipticity. However we have not considered
realistic observational noise, such as due to seeing and  possibly anisotropic
point spread functions -- thus we are assuming ideal imaging data 
in this paper. These sources of observational noise can degrade the 
signal and produce
systematic errors that must be carefully modeled to extract the
shear signal.  

The outline of our paper is as follows. In \S 2 the formalism for 
the weak lensing calculation is presented. Our N-body simulations 
and our ray-tracing method are described in \S 3, where the effective 
resolution of our results is estimated and is tested by varying the 
relevant numerical parameters. \S 4 gives predictions for the power 
spectrum and the real space 2-point correlation of the convergence 
for various cold dark matter (CDM) models. The 
scatter due to noise (primarily from the finite number and 
intrinsic ellipticity of the background galaxies) and to the finite
survey area is estimated. \S 5 contains a discussion of various 
measures of non-Gaussianity.  Results for the probability distribution 
function and its moments are presented in \S 6, where we also discuss
the survey size needed for an accurate measurement of $\Omega_{\rm m}$
given realistic number densities of source galaxies. 
A likelihood and $\chi^2$ analysis of the pdf is made in \S 7, which
also further explores robust estimators of $\Omega_{\rm m}$ based on 
the pdf. All our results are summarized in \S 8. 

\section{Theory of multiple plane lensing}

Gravitational lensing shears and magnifies the images of distant
galaxies. In this section we relate
the shear and magnification to perturbations in the
gravitational potential along the line of sight. We first analyze
the continuous case, then present a discrete approximation to it. In the 
discrete case the matter distribution between the source and observer is
taken to lie on a finite number $N$ of lens planes. The 
mapping from the image plane (the first lens plane) to the
source plane is then determined by adding the deflections 
over all the $N$ lens planes (e.g. Schneider, Ehlers \& Falco 1992).

We will work in comoving coordinates $x^i$ and conformal time $\tau$, 
in terms of 
which the perturbed Robertson-Walker metric is 
\begin{equation}
ds^2=a^2(\tau)\left\{-(1+2\phi)d\tau^2+(1-2\phi)
    \gamma_{ij}dx^idx^j\right\}.
    \label{metric}
\end{equation}
The spatial part of the background metric can be written as
\begin{eqnarray}
\gamma_{ij}dx^idx^j=d\chi^2+r^2(d\theta^2+\sin^2 \theta d\phi^2),
\nonumber \\
\nonumber \\
r(\chi)=\sin_K\chi \equiv
\left\{ \begin{array}{ll} K^{-1/2}\sin K^{1/2}\chi,\ K>0\\
\chi, \ K=0\\
(-K)^{-1/2}\sinh (-K)^{1/2}\chi,\ K<0\\
\end{array}
\right.
\label{rchi}
\end{eqnarray}
where $K$ is the curvature term which
can be expressed using the present density
parameter $\Omega_0$ 
and the present
Hubble parameter $H_0$ as $K=(\Omega_0-1)H_0^2=-\Omega_KH_0^2$.
Here $a(\tau)$ is the scale factor expressed in terms of conformal
time and $r(\chi)$ is the comoving angular diameter distance. The 
evolution of $a(\tau)$ is determined by
the Friedmann equation, $da/d\tau=H_0(\Omega_{\rm m}a+
\Omega_{\Lambda } a^4 + \Omega_Ka^2)^{1/2}$,
where $\Omega_{\rm m}, \Omega_{\Lambda }, \Omega_K$ are the densities
of matter, vacuum energy (cosmological constant) and curvature, respectively,
in units of the critical density. In addition the 
null geodesics satisfy to lowest order the relation $d\tau=d\chi$.
The density parameter $\Omega_0$ can have contributions from 
$\Omega_{\rm m}$ or $\Omega_{\Lambda}$, $\Omega_0=\Omega_{\rm m}+
\Omega_{\Lambda}$.
The advantage of using the conformal time $\tau$ is that the metric
becomes conformally Euclidean ($K=0$), 3-sphere ($K>0$) or
3-hyperboloid ($K<0$), leading to a simple geometrical description
of light propagation.
Note that we have adopted units such that $c=1$. 

The change in a photon's
direction as it propagates
is governed by the space part of the geodesic equation, which,
applied to the metric of equation \ref{metric}, gives
$d {\vec \alpha}=-2\vec{\nabla}_{\perp} \phi \ d\chi$, where $\phi$ is 
the gravitational potential, $d {\vec \alpha}$ the change in photon 
propagation direction and the symbol $\vec \nabla_\perp$ 
denotes the transverse derivative.
An individual deflection by $d \vec \alpha$
at $\chi'$ leads to a transverse excursion at $\chi$ given by
$d \vec x_\perp(\chi)=r(\chi-\chi')d \vec \alpha(\chi')$.
The final position at $\chi$ is given by the unperturbed position  
$r(\chi)\vec \theta(\chi=0)$ plus the integral along the photon 
trajectory of the individual deflections,
\begin{equation}
\vec x_{\perp}(\chi)=-2\int_0^\chi r(\chi-\chi')\
\vec\nabla_\perp\phi[\chi', \vec x_{\perp}(\chi')]\
d\chi'+r(\chi)\, \vec \theta(\chi=0) .
\label{xperp}
\end{equation}
Note that the gravitational potential has to be computed along the 
actual perturbed path of the photon, which we denote with radial 
position $\chi'$ and transverse position $\vec x_{\perp}(\chi')$.
The angular direction at $\chi$ is given by 
\begin{equation}
\vec \theta(\chi)={-2 \over r({\chi}) } \int_0^{\chi}r(\chi-\chi')\
\vec \nabla_\perp \phi[\chi',r(\chi')\vec \theta(\chi')]\ d\chi'+
\vec \theta(\chi=0).
\label{dtheta}
\end{equation}
From the above we define the shear matrix as
\begin{eqnarray}
\Phi_{ij} &\equiv & {\partial \theta_i(\chi) \over \partial \theta_j(0)}
-\delta_{ij}
=- 2 \int_0^{\chi}
g(\chi',\chi)\ \nabla_i \nabla_j 
\phi[\chi',r(\chi')\vec \theta(\chi')]\ d\chi'\ ,\nonumber \\
g(\chi',\chi)& =& {r(\chi')r(\chi-\chi') \over r(\chi) }\ , 
\end{eqnarray}
where the indices $i,j$ take values $1,2$ and denote components
on the sky. 

The convergence of the mapping is defined as $2\kappa=\Phi_{11}+\Phi_{22}$.
The two components of the shear $\bi{\gamma}=\gamma_1+i\gamma_2$ 
are similarly defined by:
$2\gamma_1=\Phi_{11}-\Phi_{22}\ ; \ \gamma_2=\Phi_{12}\, . $
To obtain $\kappa$ one needs to compute $\int_0^{\chi}
g(\nabla_{11}+\nabla_{22})\phi\, d \chi'$ along the photon trajectory. This 
integral can be simplified by using the projected density distribution.
The relation between the gravitational potential $\phi$
and density perturbation $\delta$ is given by Poisson's equation
\begin{equation}
\nabla^2 \phi={ 3  H_0^2 \over 2 }\ \Omega_{\rm m}\ {\delta \over a}\ .
\end{equation}
Using this equation the expression for $\kappa$ in terms of $\Phi_{11}$ 
and $\Phi_{22}$ becomes
\begin{equation}
\kappa=\int_0^{\chi}g
[\nabla^2-\nabla_{\chi}^2]\phi\ d \chi'=
\int_0^{\chi}{ 3  H_0^2 \over 2 }\ \Omega_{\rm m}\ g {\delta \over a}\ 
d \chi'-\int_0^{\chi}g\nabla_{\chi}^2\phi\ 
d \chi'.
\label{deco}
\end{equation}
The second term in equation (\ref{deco}) can be integrated by parts,
\begin{equation}
\int_0^{\chi}g\nabla_{\chi}^2\phi\ d\chi'=
g\nabla_{\chi}\phi |_0^{\chi}-\int_0^{\chi}[
g \dot{\nabla}_{\chi}\phi+g'\nabla_{\chi}\phi]\ d\chi'.
\end{equation}
The prime and dot denote derivatives with respect to the radial coordinate
$\chi$ and conformal time $\tau$, respectively. Note that we are 
integrating along photon geodesics and hence $d/d\chi=\partial/
\partial \tau + \partial/ \partial \chi$.
The boundary terms vanish because $g=0$ at $\chi'=0$ and at $\chi'=\chi$.
The time derivative of $\nabla_{\chi}$ typically 
changes by order unity over the 
Hubble time and the term is comparable to $g'\nabla_{\chi} \sim 
\nabla_{\chi}$. These two terms have to be compared to 
$ g \nabla^2_{\perp}\sim \chi 
\nabla^2_{\perp}$. On small angular scales the dominant contribution 
to the deflections comes from scales much smaller than  
$\chi$, which is comparable to the Hubble length. 
Hence $\nabla_{\perp}^2 \gg \nabla_{\chi}/\chi$ 
and the second term in equation (\ref{deco})
can be ignored compared to the first term everywhere except on the largest
angular scales. The convergence can therefore be expressed as a
projection of the density perturbation $\delta$
\begin{equation}
\kappa={ 3  H_0^2 \over 2 }\ \Omega_{\rm m}\ \int_0^{\chi}g\ {\delta \over a}
\ d \chi'.
\label{Kproj}
\end{equation}
We will discuss other components of $\Phi$ in the following subsection.

The power spectrum of the convergence on angular wavenumber $l$ can be 
expressed in terms of the
density power spectrum using equation (\ref{Kproj}). 
The result for the dimensionless power which
gives the contribution to the variance per log interval in $l$ is 
(Jain \& Seljak 1997)
\begin{equation}
\sigma^2(l)\ \equiv \ 2\pi l^2 P(l) = 36 \pi^2 \Omega_m^2 l^2 \int_0^\chi
\frac{g^2}{r^2} \ \frac{P_\delta(k,\chi')}{a^2} \ d\chi',
\label{kappapower}
\end{equation}	
where $P_\delta(k,\chi)$ is the power spectrum of the density at 
wavenumber $k=l/r$. 

\subsection{Discrete lensing approximation}

We may discretize the expressions above by dividing the radial 
interval between source and observer into $N$ planes separated 
by comoving distance $\Delta \chi$. The angular position of a ray
at the $n$-th plane is given by equation \ref{dtheta} as
\begin{eqnarray}
{\vec \theta}_n&=&\sum_{m=1}^{n-1} {r(\chi_n-\chi_m) \over r(\chi_n)}
{\vec \alpha}_m+{\vec \theta}_1
\nonumber \\
{\vec \alpha}_m&=&-2{\vec \nabla}_{\perp}
\phi[\chi_m,r(\chi_m){\vec \theta}_m] \Delta \chi \, .
\end{eqnarray}
We have assumed that the spacing between planes is sufficiently small
that only the lowest order terms contribute. 
In analogy with the continuous case 
we can define the Jacobian of the mapping $\Phi_n$, which describes the 
deflections in the $n$-th plane relative to the image plane
\begin{equation}
\Phi_n= {\partial {\vec \theta}_n \over \partial {\vec \theta}_1}.
\end{equation}
The shear tensor in each plane $U_m$ is defined as 
\begin{equation}
{\partial {\vec \alpha}_m \over \partial {\vec \theta}_m}=r(\chi_m)
{\partial  {\vec \alpha}_m
\over \partial {\vec x}_m}=
-2r(\chi_m)\vec{\nabla}_{\perp}{\vec\nabla}_{\perp}
\phi(r(\chi_m){\vec \theta}_m) \Delta \chi \equiv r(\chi_m) U_m \ .
\label{u}
\end{equation}
The distortion tensor can now be written as (Schneider, Ehlers \& Falco 1992;
Seitz, Schneider \& Ehlers 1994)
\begin{eqnarray}
\Phi_n& = &{\partial {\vec \theta}_n \over \partial {\vec \theta}_1}=
 I + \sum_{m=1}^{n-1} {r(\chi_n-\chi_m) \over r(\chi_n)}
{\partial {\vec \alpha}_m \over \partial {\vec \theta}_m}
{\partial {\vec \theta}_m \over \partial {\vec \theta}_1} \nonumber \\
& = & I + \sum_{m=1}^{n-1} g_{mn} U_m \Phi_m,
\label{Amult}
\end{eqnarray}
where $g_{mn}=r(\chi_n-\chi_m)r(\chi_m)/r(\chi_n)$, 
and $I$ denotes the identity matrix. 
Note that although the matrices $U_m$ are symmetric for all $m$, the
matrices $\Phi_m$ are not symmetric for $m>2$. The symmetry is destroyed
because the matrix products in equation
\ref{Amult} are not symmetric even if their component matrices 
are. The recursion relation in equation \ref{Amult} 
is used by our ray-tracing code to propagate
the distortion tensor. The most time-consuming computational step 
is the evaluation of the $U_m$ matrices at each photon position and 
on each lens plane.

In a single lens plane the 2 by 2 matrix $U_m$ at a given angular 
position is symmetric
and is given by the integral of the second derivatives of the gravitational 
potential $\phi$ along the radial extent of the $m$-th region  
(equation \ref{u}). 
Following the discussion for the continuous case this can be approximated 
in terms of the projected surface density 
$\Sigma=\int_{\chi_{m-1}}^{\chi_m}\ \delta\ d\chi'$. We define
\begin{equation}
\Kappa=\frac{1}{2}\left(U_{11}+U_{22}\right)={ 3  H_0^2 \over 2 } \, 
\Omega_{\rm m}\ g\, {\Sigma  \over a}\ .
\label{sigma}
\end{equation}
The other two components of the symmetric tensor are 
\begin{eqnarray}
\Gamma_1&=&\frac{1}{2}\left(U_{11}-U_{22}\right)\ , \nonumber \\
\Gamma_2&=&U_{12}\ .
\label{Umatrix}
\end{eqnarray}
They can be obtained from $\Kappa$
using the relation between these
quantities in Fourier space (which follows from the definition of $U$
in equation \ref{u}):
\begin{eqnarray}
\hat\Gamma_1(\vec l)&=&\frac{l_1^2-l_2^2}{l^2}\hat\Kappa(\vec l)\ ,\nonumber \\
\hat\Gamma_2(\vec l)&=&\frac{2l_1 l_2}{l^2}\hat\Kappa(\vec l)\ .
\label{gammaft}
\end{eqnarray}
We have used the following definition of the 2-dimensional Fourier 
transform of $X=\Kappa,\Gamma_1, \Gamma_2$
\begin{equation}
X(\vec \theta)=\int d^2l e^{i {\vec l}\cdot \vec \theta}\hat X({\vec l})\ .
\label{ft}
\end{equation}

Finally, the shear matrix $U_m$ is computed at the perturbed photon 
positions ${\vec \theta}_m={\vec \theta}_1+\delta {\vec \theta}_m$. 
The perturbations $\delta {\vec \theta}_m$ can be computed from the Jacobian
matrix $\Phi_m$ using the Fourier space relations
\begin{eqnarray}
\delta \hat{\theta}_1=-i\hat{\Phi}_{11}/l_1 \nonumber \\
\delta \hat{\theta}_2=-i\hat{\Phi}_{22}/l_2,
\label{thetaper}
\end{eqnarray}
where we have dropped the index $m$ for clarity. One could also
express the above relations in terms of off-diagonal elements of $\Phi$.

\subsection{Magnification, Shear and Rotation of Galaxy Images}

To obtain the ellipticity induced by gravitational lensing
we consider the action of the distortion
matrix $\Phi$ on a circular image. A 
general matrix can be decomposed into a symmetric and a rotational 
matrix (Schneider, Ehlers \& Falco 1992) 
\begin{equation}
\Phi= \left(
\begin{array}{cc}
\Phi_{11} & \Phi_{12} \\
\Phi_{21} & \Phi_{22} \\
\end{array}
\right) \ =\left(
\begin{array}{cc}
\cos \phi & \sin \phi \\
-\sin \phi & \cos \phi \\
\end{array}
\right) \
\left(
\begin{array}{cc}
1-\kappa-\gamma_1 & -\gamma_2 \\
-\gamma_2 & 1-\kappa+\gamma_1 \\
\end{array}
\right) \ .
\label{Amatrix}
\end{equation}
In the symmetric case $\phi=0$ and
the image of a circular source is stretched into an ellipse, 
with the stretch factors of the two axes given 
by $(1-\kappa \pm \gamma)^{-1}$, where $\gamma^2=\gamma_1^2+\gamma_2^2$.
The direction of major axis of the ellipse 
is $\beta=\tan^{-1}(\gamma_2/\gamma_1)/2$.
The magnification of the image is given by $1/\det(\Phi)$.

If the distortion matrix $\Phi$ is not symmetric there is also 
an overall rotation of the image by angle $\phi$. 
Assuming this term is small we can expand the rotation matrix
to obtain 
\begin{equation}
\phi \approx {\Phi_{12}-\Phi_{21} \over 2(1-\kappa)}.
\end{equation}
Image rotation by $\phi$ leaves the magnification unchanged to lowest order,
but modifies the two shear components: $\gamma_1'=\gamma_1+2\gamma_2\phi$
and $\gamma_2'=\gamma_2-2\gamma_1\phi$. In general, the lowest order
contributions to the rotation angle $\phi$ are quadratic in shear and are
significantly smaller than the first order contributions to
$\kappa$ and $\gamma$. The only situation in which rotation can be 
important is if two critical images that are 
well separated in redshift are superimposed; this
is rather unlikely in our universe. We verify this below by measuring the 
rotational component of $\Phi$ in our simulations -- it is shown
in figure \ref{figantisym} and discussed in Section 4. 

For individual galaxy images the gravitational stretching cannot
be separated from the intrinsic ellipticity. However, by averaging
the ellipticity of all images seen near a given direction, a
smoothed  version of the gravitational component can be measured. 
To compute the distortion in this case we need to average the 
distortion matrices over the smoothing window and over the redshift 
distribution of the sources. Let us assume we know the probability 
$W_m$ that a given galaxy lies on the $m$-th plane, where 
$\sum_m W_m=1$. The average distortion before angular smoothing
is then 
\begin{equation}
\Phi=\sum_m W_m \Phi_m \ .
\label{weight}
\end{equation} 
Ray-tracing must therefore be performed up to the 
most distant plane where  $W_m$ is not negligible.

\section{Ray-tracing through N-body simulations}

\subsection{Ray-tracing procedures}

The previous section contains the formulae for multiple plane
lensing that we use in our ray tracing simulations. The
key relations are equations
(\ref{thetaper}) for the position of a given photon ${\vec \theta}_n$ 
and equations (\ref{Amult}-\ref{gammaft}) for the Jacobian matrix $\Phi_n$ 
at this position. Aside from the distance factor $g_{mn}$,
the main inputs into the recursion relations for ${\vec \theta}_n$ 
and $\Phi_n$ are the shear matrices $U_m$ and previous $\Phi_m$'s, 
all of which need to be stored. 

The ray tracing algorithm consists of three parts: constructing
the dark matter lens planes, computing the
shear matrix on each plane, and using these to evolve the photon
trajectory from the observer to the source. The details involved
at each step are as follows. 

\noindent 1. The dark matter distribution between the source and 
observer is projected onto $20-30$ equally spaced lens planes, depending
on the size of simulation box and the redshift of the sources.
The particle dump closest in redshift to each plane is used for
this projection. The projected field is a square of predetermined 
angular size -- usually the angular size of the simulation box at 
the source distance. The orientation of the projection as well as 
the location of the origin are chosen at random. This minimizes 
repetition of structures from the simulation at the same projected 
position. The particle positions on each plane are interpolated 
onto a grid of size $2048^2$ for the P$^3$M runs, and smaller sizes
for the PM runs (the P$^3$M and PM simulations are described
in the next sub-section). The dimensionless density $\Kappa$ used in
the lensing equations is given by the projected particle density
times a redshift-dependent constant (see equation \ref{sigma}). 

\noindent 2. On each plane $U$ is computed on a grid. 
This is done by Fourier transforming the projected density $\Kappa$ 
and using equations (\ref{gammaft}) to obtain the shear. The inverse
Fourier transform is then used to return to real space, and the
shear matrix $U$ is given by linear combinations of $\Kappa$, 
$\Gamma_1$ and $\Gamma_2$ obtained from equations (\ref{sigma}) 
and (\ref{Umatrix}). Periodic boundary conditions are assumed when
carrying out these operations, hence we use zero-padding to minimize
the spurious effects introduced. Since the final convergence effectively
depends only on a projection of the density along the photon path
(equation \ref{Kproj}) and the relative location of different planes
is in any case random transverse to the line of sight, this 
mistreatment of the effects of the matter outside our grid has very little
influence on our results. 

\noindent 3. The photons are started on a regular grid on the first
lens plane. Perturbations along the line of sight distort this grid
and are computed using equation 
(\ref{thetaper}). Once we have the photon positions, we
interpolate the shear matrix $U$ onto them and 
solve equation \ref{Amult} 
for the Jacobian matrix $\Phi_n$ of the mapping from the $n$-th lens plane
to the first plane, using previously stored $\Phi_m$ and $U_m$. 

\noindent 4. Solving the recursion relations up to the source
plane yields the Jacobian matrix $\Phi_n$ at these positions. 
Finally, we may sum over the 
redshift distribution of sources using equation \ref{weight}. 
Note that the ray tracing is done
backwards from the observer to the source, thus ensuring that all
the photons reach the observer. The first lens plane is the image
plane and has the unperturbed photon positions. 

The use of FFT's for computing ${\vec \theta}$ and $U$ simplifies 
and speeds up the numerical implementation. Because
the CPU required for an $M^2$ FFT scales as $M^2 \log M^2$, 
it is possible to use sufficiently large grids to maintain the 
intrinsic resolution of the N-body simulation (see below). A direct 
summation would be prohibitively time 
consuming. An alternative Poisson-solver of similar resolution
would be a tree-code as
used by Wambsganss et al (1995) and Wambsganss, Cen \& Ostriker (1998). 

We found that a substantial fraction of the overall time in a ray-tracing
run was taken up by reading in the $\Sigma$ grids. The 2-dimensional 
FFT's needed for computing $\vec\alpha$ and $U$ take the bulk of the 
CPU time. CPU constraints turn out, however, to be negligible: a run using
grids of size $1024^2$ can be completed in 20-30 minutes on a DEC-alpha
workstation provided it has enough memory. Memory is a problem with 
such large grids because the recursion relations require the storage
of the $\Phi_n$ matrices at all the lens planes. This could be avoided 
if one used the recursion relation based on previous two planes only
(Seitz, Schneider \& Ehlers 1994), although for simplicity 
we have not implemented that. Thus for the 4-component
matrix of real*4 numbers for the $\Phi_n$'s, computed on $1024^2$ 
photon positions on 20 lens planes, one needs 
$4\times1024^2\times20\times4$ bytes, 
or about 320 Megabytes of memory. A $2048^2$ grid with a spatial resolution
twice as good would require 4 times the memory, i.e. more than 
1.2 Gigabytes -- this is the grid size used for our high resolution runs. 

\begin{center}
Table 1: Parameters of the N-body simulations\\
\bigskip
\begin{tabular}{|r|llll|}
\hline
Simulation & SCDM & $\tau$CDM & $\Lambda$CDM & OCDM\\
\hline
$N_{\rm par}$ & 256$^3$ & 256$^3$ & 256$^3$ & 256$^3$ \\
$l_{\rm soft}[h^{-1}$ kpc]& 36 & 36 &30 & 30 \\
$\Gamma$ & 0.5 & 0.21 & 0.21 & 0.21\\
$L_{\rm box}[h^{-1}$Mpc] & 85 & 85 & 141 & 141 \\
$\Omega_0$ & 1.0 & 1.0 & 0.3 & 0.3 \\
$\Lambda_0$ & 0.0 & 0.0 & 0.7 & 0.0 \\
$H_0$ [km/s/Mpc] & 50 & 50 & 70 & 70 \\
$\sigma_8$ & 0.6 & 0.6 & 0.9 & 0.85 \\
$m_p[h^{-1} 10^{10} M_{\odot}]$ & $1.0$ & $1.0$ &
$1.4$ & $1.4$\\
\hline
\end{tabular}
\end{center}


\subsection{The N-body simulations}

Two sets of N-body simulations are used in this paper. Most of our
results are based on a set of high resolution,  adaptive
particle-particle/particle-mesh (AP$^3$M) simulations, but we also
use a set of particle-mesh (PM) simulations to explore issues related
to numerical resolution.
PM simulations solve Poisson's equation on a grid in order 
to advance the particle distribution. The grid spacing is the key
factor limiting the numerical resolution; we use grids
of sizes between $512^3$ and $128^3$ with a particle number typically equal to 
one eighth the number of grid sites. AP$^3$M simulations supplement a grid
calculation of the long-range component of the gravitational force
with a short range correction computed either by a direct sum over
neighboring particles, or, in highly clustered regions, by combining a
calculation on a localized refinement mesh with a direct sum over
a smaller number of much closer neighbors. These techniques ensure
that the force law is effectively Newtonian down to much smaller
separations than can be achieved with a PM code.

Our high-resolution simulations used a parallel adaptive AP$^3$M 
code (Couchman, Thomas \& Pearce 1995; Pearce \& Couchman 1997) kindly
made available by the Virgo Supercomputing Consortium (e.g. Jenkins 
et al 1998). They followed $256^3$ particles using a force law with
softening length $l_{\rm soft}\simeq 30\ h^{-1}$kpc at $z=0$ (the
force is $\sim 1/2$ its $1/r^2$ value at one softening length and is
almost exactly Newtonian beyond two softening lengths). $l_{\rm
soft}$ was kept constant in physical coordinates over the redshift
range of interest to us here. The simulations were carried using 128
or 256 processors on CRAY T3D machines at the Edinburgh Parallel 
Computer Centre and at the Garching Computer Center of the Max-Planck
Society. The particle distribution was evolved from a starting redshift
of $z=50$ to $z=0$. Table 1 gives the essential cosmological and numerical 
parameters defining these simulations. They have previously been 
used for studies of strong lensing by Bartelmann et al (1998), for 
studies of dark matter clustering by Jenkins et al (1998), and for 
studies of the relation between galaxy formation and galaxy clustering
by Kauffmann et al (1998a,b) and Diaferio et al (1998).

\begin{figure}[t!]
\vspace*{9.5cm}
\caption{Effective angular resolution versus redshift of the lens plane. 
The solid ($\Omega_{\rm m}=1$) and dotted ($\Omega_{\rm m}=0.3$) curves show the scale 
corresponding to the force softening length of the N-body simulations. 
The short- ($\Omega_{\rm m}=1$) and long- ($\Omega_{\rm m}=0.3$) dashed lines show
the scale given by twice the grid spacing used in the ray tracing. 
The lensing efficiency is peaked between $z=0.3-0.5$ and falls off
at lower and higher $z$. As a result, the steep rise in the force
softening scale at low-$z$ does not strongly affect the final 
resolution.
}
\includegraphics{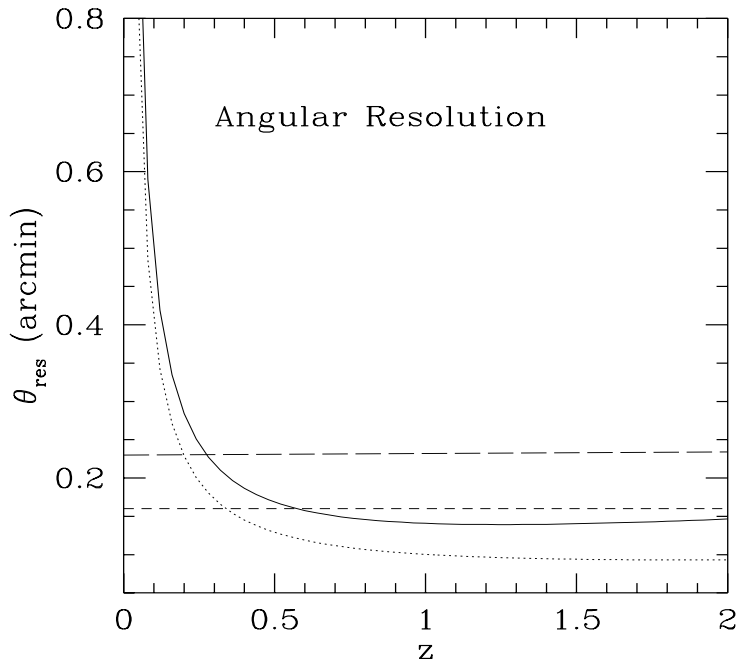}
\label{figres}
\end{figure}

\begin{figure}[t!]
\vspace*{9cm}
\caption{The effect of varying the number of particles selected from
the N-body simulation on the power spectrum of $\gamma$. The 
wavenumber $l$ (in inverse radians) is plotted up to the Nyquist frequency
of the ray tracing grid. 
The curves shown by the solid and dotted
lines are for all the particles, $N_{\rm particle}=256^3$, and one eighth
the particles, respectively. The dotted curves have greater
power at high $l$ since the white noise contribution is larger
with fewer particles.
}
\includegraphics{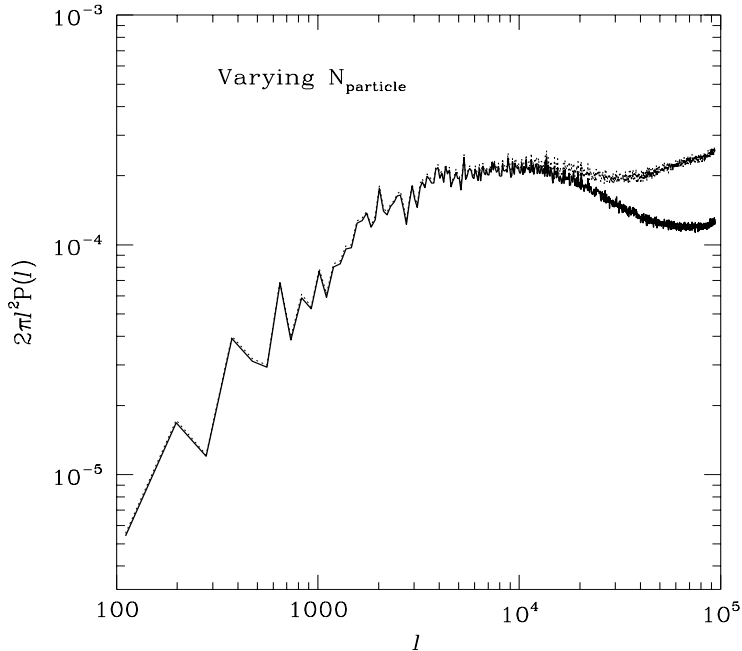}
\label{figpart}
\end{figure}

\begin{figure}[p!]
\vspace*{16.5cm}
\caption{The effect of the interpolation scheme and grid size 
on the convergence power spectrum (upper panel) and on the 2nd, 3rd 
moments of the mean convergence in square fields on the sky (lower panels). 
In the upper panel the two solid curves extending to $l=10^5$ are
for a $2048^2$ ray tracing grid, and the two dotted curves for a
$1024^2$ grid. The power spectrum in each case is plotted up to $l$ 
corresponding to the Nyquist frequency of the ray
tracing grid. For given grid size, the curve with more power at
high-$l$ used nearest-grid-point interpolation while the lower 
curves used cloud-in-cell interpolation. 
}
\includegraphics{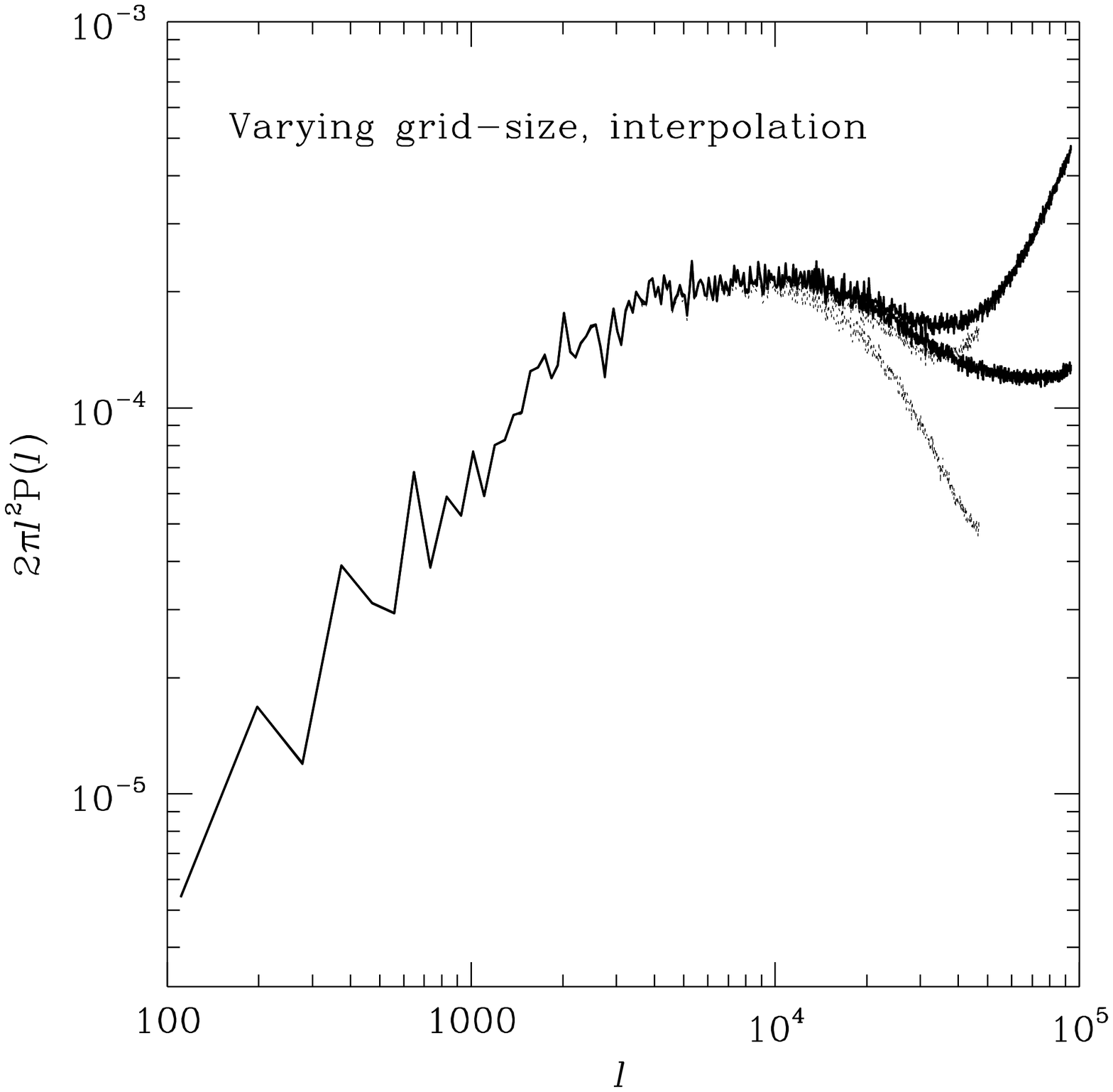}
\includegraphics{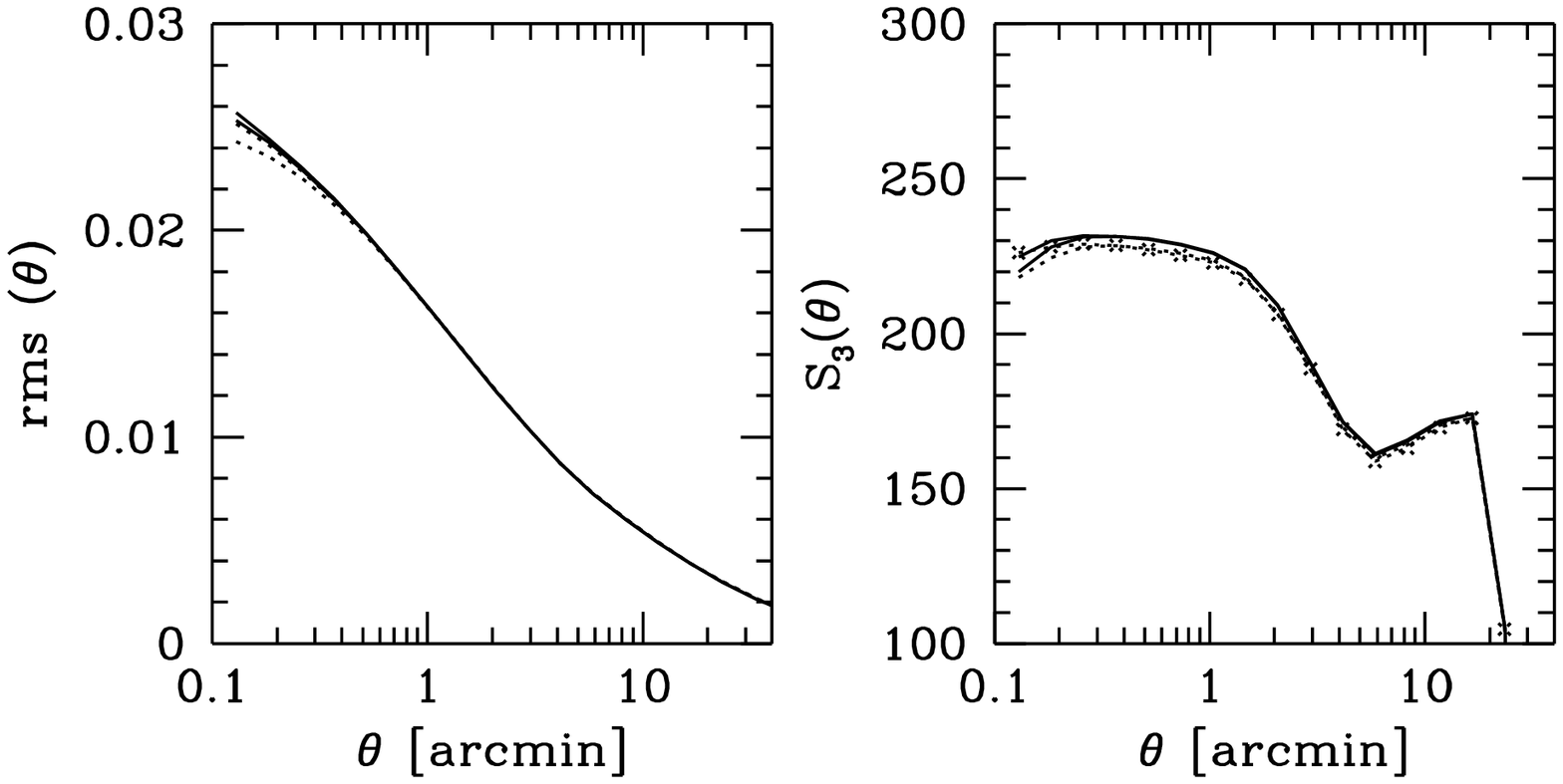}
\label{figinterp}
\end{figure}
 
\subsection{Numerical resolution issues}

There are two kinds of resolution limitation in our ray-tracing simulations. 
The first reflects the finite size and resolution of our N-body simulations,
the second our use of finite grids when computing deflection angles
and shear tensors on the lens planes. Our main consideration when 
choosing the size of the grids was to maintain the resolution 
provided by the N-body simulation. Figure \ref{figres} compares the
angular scales corresponding to the size of the ray-tracing grid, 
and to the spatial resolution of the N-body simulations. The rest
of this sub-section provides a detailed discussion of how these 
scales affect our results based on explicit numerical checks made
by varying the relevant simulation parameters. 

\noindent{\bf Force softening.} 
The spatial and mass resolution of the AP$^3$M simulations are given in
Table 1 by the parameters $l_{\rm soft}$, the effective force softening 
length, and $m_{\rm p}$, the particle mass. $l_{\rm soft}$ 
corresponds to different angular scales at 
different redshifts. A rough estimate of the angular resolution limit 
is provided by the lensing planes at $z\simeq 0.3$. The lensing contribution
peaks in the range $0.3\lsim z\lsim 0.5$ for realistic values of $\Omega_{\rm m}$ and
$z_{\rm source}$. For $\Omega_{\rm m}=1$, the angular scale corresponding to 
$l_{\rm soft}$ at $z=0.3$ is $\theta_{\rm soft}=0.18'$; for 
lower $\Omega_{\rm m}$ or higher redshift, $\theta_{\rm soft}$ is
smaller (see figure 2). The effect of softening in N-body simulations
is to prevent the formation of objects much smaller than $l_{\rm soft}$ 
and to soften central density cusps in larger objects at radii
below  $l_{\rm soft}$. Our lensing statistics will thus be
affected by softening only insofar as they are sensitive to such small
regions.

\noindent{\bf Mass resolution.} 
The effect of N-body discreteness on our derived lensing statistics can 
be quantified if we assume that the particle distribution in the 
simulation can be modeled as a random Poisson sampling from an
effectively continuous density field defined by the true mass
distribution in the region simulated (i.e. the mass distribution the
simulation would have produced if it had been carried out using a
very large number of particles, resulting in very fine mass resolution).
We limit the analysis here to second order statistics although it is
straightforwardly generalized to higher order moments. From equation
(\ref{Kproj}) it is easy to show that Poisson sampling introduces an
additional variance in the convergence along each line of sight given by
\begin{equation}
{\rm Var}_P(\kappa) = \big( {3H_0^2\over 4}\Omega_m\big)^2
  {m_p\over \omega\bar{\rho}} \int_0^\chi {g^2\over a^2 {dV\over
  d\chi'}}(1+\delta) d\chi'\, ,
\end{equation}
where $m_p$ is the mass of a particle in the N-body simulation,
$\omega$ is the effective solid angle over which averaging is carried
out to evaluate $\kappa$ and $V(\chi')$ is the comoving volume per
steradian out to distance $\chi'$. The Poisson sampling along
different lines of sight is independent, so it produces an additional
white noise contribution to the power on all scales which, 
averaged over sky patches of solid angle $\omega$, has variance
\begin{equation}
\sigma_P^2(\kappa) = \big( {3H_0^2\over 4}\Omega_m\big)^2
  {m_p\over \omega\bar{\rho}} \int_0^\chi {g^2\over a^2 {dV\over
  d\chi'}} d\chi'\, .
\label{Poisvar}
\end{equation}
Note that this does not depend on the fluctuations, $\delta({\bf x})$,
and as a result the power spectrum of the discrete distribution is
just the sum of the intrinsic power spectrum of the continuous field
and a white noise power discreteness term.
In this Poisson sampling model the simulated power spectra are thus
easily corrected for discreteness. The remaining effects are
then added noise in power spectrum estimates on scales where 
discreteness dominates, together with systematic
effects on very small scales due to inadequacies of the Poisson
sampling model.

In our high resolution simulations discreteness effects are more
significant in the Einstein-de Sitter models. This is because the
simulations are normalized to have the same cluster abundance by 
mass, and as a result have similar convergence power spectra on scales
of a few arcminutes. The white noise power level from the above equations 
is roughly proportional to $m_p\, (\Omega_mH_0^2)^2$ and so is 
higher than in the low density models.
Figure \ref{figpart} shows the effect on the ellipticity power spectrum 
of varying the 
number of particles selected from the N-body simulations. The dotted curve, 
which has greater power at high $l$, shows the power spectrum for
a run in which one eighth the particles chosen at random were removed. 
While the effect of discreteness and finite grids (discussed below) can
be modeled and corrected for in the case of the power spectrum, we have 
suppressed these effects by smoothing the projected density distribution, 
with smoothing length typically of order $\theta_{\rm soft}$. 
The form of the smoothing function used 
is $(\theta^2+\theta_{\rm smooth}^2)^{-3/2}$. 
Its 2-dimensional Fourier transform is an exponential, therefore the
smoothing convolution is conveniently performed in Fourier space.

\noindent{\bf Variation with lens redshift.}
The effects of force softening and finite particle number are both
very pronounced for the low redshift lens planes. The reason is that
the side length of the square subtended by a beam of fixed angular size
decreases with decreasing redshift. It is
less than 1/10th the side length of the N-body simulation box
at $z\lsim 0.1$, so that fewer than 1/100th of the particles
contribute to the projected density. The force softening is also larger
in angular units compared to its value at higher redshift. Because
the lowest redshift planes can contribute significantly to the
convergence, Poisson sampling effects and smoothing in their mass
distribution are not always negligible. This is apparent in the
equation for the Poisson sampling variance given above where the
integrand is significantly more strongly weighted to low redshift than
the integrand of the corresponding integral for the convergence
(cf equations (\ref{Poisvar}) and (\ref{Kproj})).

\noindent{\bf Grid effects.}
The effect of the 2-dimensional grids used in the ray tracing
enters at several points. First the density is interpolated onto
the grid and used to compute the shear tensor and then the shear tensor 
and deflection angle are interpolated from the grid to the ray 
positions. 
In both steps we expect that numerical 
smoothing would occur on scales of order twice the grid spacing. 
We therefore assign the grid resolution to be twice the grid spacing
giving $\theta_{\rm grid}=0.16'$ for the $\Omega_{\rm m}=1$ models, and 
to $\theta_{\rm grid}=0.23'$ for the open and $\Lambda$ models. In our 
high-resolution ray tracing runs $\theta_{\rm grid}$ is kept constant at 
all redshifts. 
Figure \ref{figinterp} shows 
the effects of the interpolation scheme and grid size on 
the power spectrum of the ellipticity. 
These effects are significant for the power spectrum at high-$l$
as expected. At high-$l$ for the $2048^2$ grids the effect 
of finite particle number, shown 
in figure \ref{figpart} and discussed above, dominates the
suppression due to grid effects and produces an $l^2$ tail in
the dimensionless power. The real space moments
shown in the lower panels of figure \ref{figinterp} are unaffected
by these high-$l$ effects. 
The real space moments on small scales, measured using 
top-hat windows, are dominated by longer-wave modes since
the dimensionless power peaks between $l=10^3-10^4$.

\noindent{\bf Small scale resolution.}
At the peak redshift of the lensing contribution, $\theta_{\rm soft}$
and $\theta_{\rm grid}$ are comparable and of order $0.2'$. However, since the 
lens efficiency is not very sharply peaked, effects at other 
redshifts also enter. 
For distant lens planes with $z\simeq z_{\rm source}$ the limiting resolution
is dominated by $\theta_{\rm grid} \sim 0.2'$. At the other extreme, 
$z\simeq 0$ force softening dominates and gives $\theta_{\rm soft}\sim 0.4'$. 
Figure \ref{figres} shows the angular resolution scale due to force
softening and finite grid spacing as a function of the redshift of the lens
plane. In summary, the above estimates and tests suggest that in our 
high-resolution runs 
the ray tracing results are reliable on scales greater than 2-4 grid
spacings, corresponding to $\sim 0.2'-0.4'$ or $l$-values below
1 or $2\times 10^4$. The lower end of the angular
range can be reached provided effects due to discreteness and
softening are corrected for carefully.

\noindent{\bf Finite box-size.}
On large scales the finite box-size of the N-body simulations sets
the upper limit on the angular scales available. For the high-resolution
runs with the smallest box size, the angular size at $z=1$ is 2.8$^\circ$. 
Thus on scales comparable to $1^\circ$, only a few modes contribute to the
power spectrum, leading to large fluctuations across different realizations. 
We therefore use several realizations of the ray tracing and average the
power spectra and other statistics to obtain accurate measurements on 
large scales. 
The real space moments on large scales also show large fluctuations. Like 
the power spectrum, the rms is a measure of the second moment, but since 
it is given by an integral over the power spectrum at low-$l$ it is 
systematically suppressed relative to the value obtained from larger 
box-sizes. The skewness and higher moments
show significantly more fluctuations than the second moment, and 
require many more realizations to obtain robust estimates. 
The dominant fluctuations
come from low redshift lens planes, whose transverse size
is $\lsim 10\ h^{-1}$ Mpc. The large scale fluctuations should be
present in observational data also -- estimates of their scatter 
set the sample size needed for accurately measuring the desired statistics. 

Since we are projecting the density through the entire box, finite box
size also implies finite number of projections, typically 20-30 up to 
$z \sim 1$. We verified the discretness effects of this approximation 
by comparing the results between $PM$ simulations of different box
sizes, typically 64$h^{-1}$Mpc and 128$h^{-1}$Mpc. We found good agreement
in the regime where the smaller resolution of the larger box simulation
was not important. Projecting over 100$h^{-1}$Mpc is not a significant
source of error and only with significantly larger projections, 200$h^{-1}$Mpc
and above, can discretness effects become important, at least if the 
majority of galaxies are at $z\sim 1$ or larger.

\noindent{\bf Multiple ray tracing realizations.}
We generated multiple realizations of shear and magnification maps
via ray tracing through a given N-body simulation. Only a fraction
of the projected area from a simulation box at a given redshift is used 
to make a lens plane,
and further the projection is made along different directions and the 
projected area rotated. By exploiting this freedom,
between 5-10 nearly independent realizations of the ray tracing
can be generated per N-body simulation. We also used P$^3$M
simulations with larger box sizes performed by the Virgo consortium, 
and several PM N-body simulations with box sizes $\gsim 200 h^{-1}$ Mpc. 
With these multiple realizations of both the ray tracing and the N-body
simulations the power spectrum and real space moments on large scales
can be measured. By comparing results from small and large box-sizes
we can also check if missing modes in the smaller
boxes lead to incorrect results on smaller scales through nonlinear
mode coupling. Likewise, the small-scale resolution limit of the larger boxes 
could be probed by comparing with the smaller boxes. 

\section{Results on the power spectrum}

\begin{figure}[p]
\vspace*{17.5cm}
\caption{The magnification pattern on a 1$^\circ$ field for an
Einstein-de Sitter (upper panel) and open (lower panel) model 
with $\Omega_{\rm m}=0.3$. 
The typical value of the magnification in these fields is $5\%$. 
}
\includegraphics{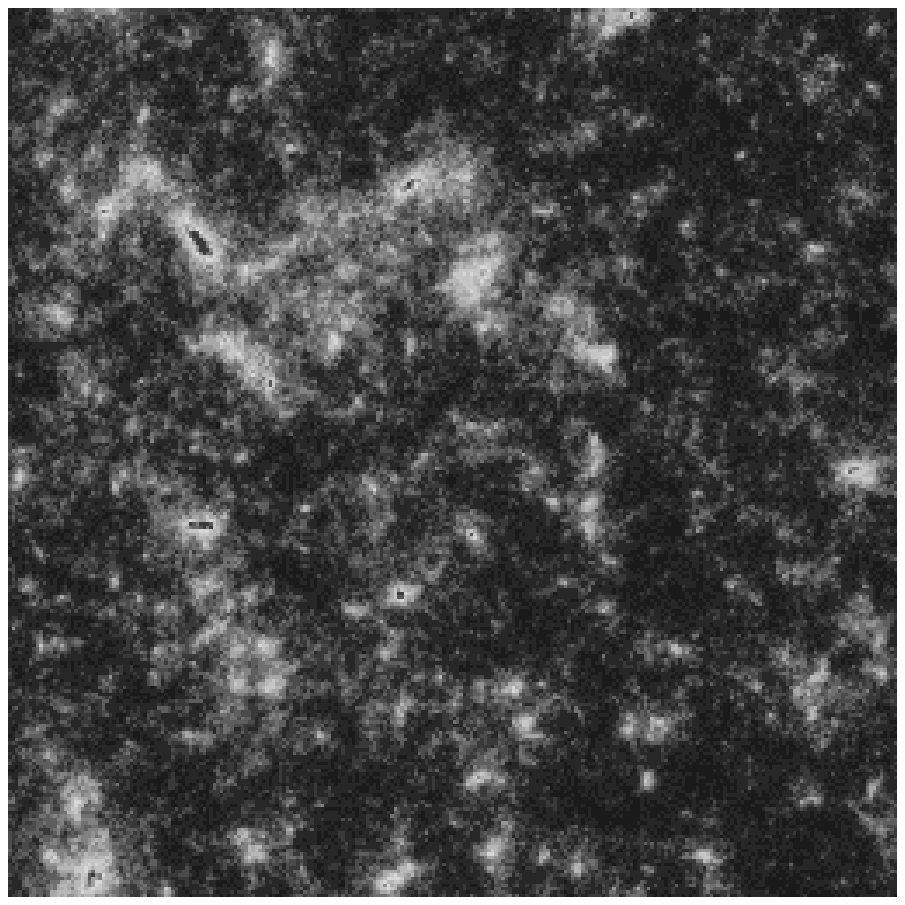}
\includegraphics{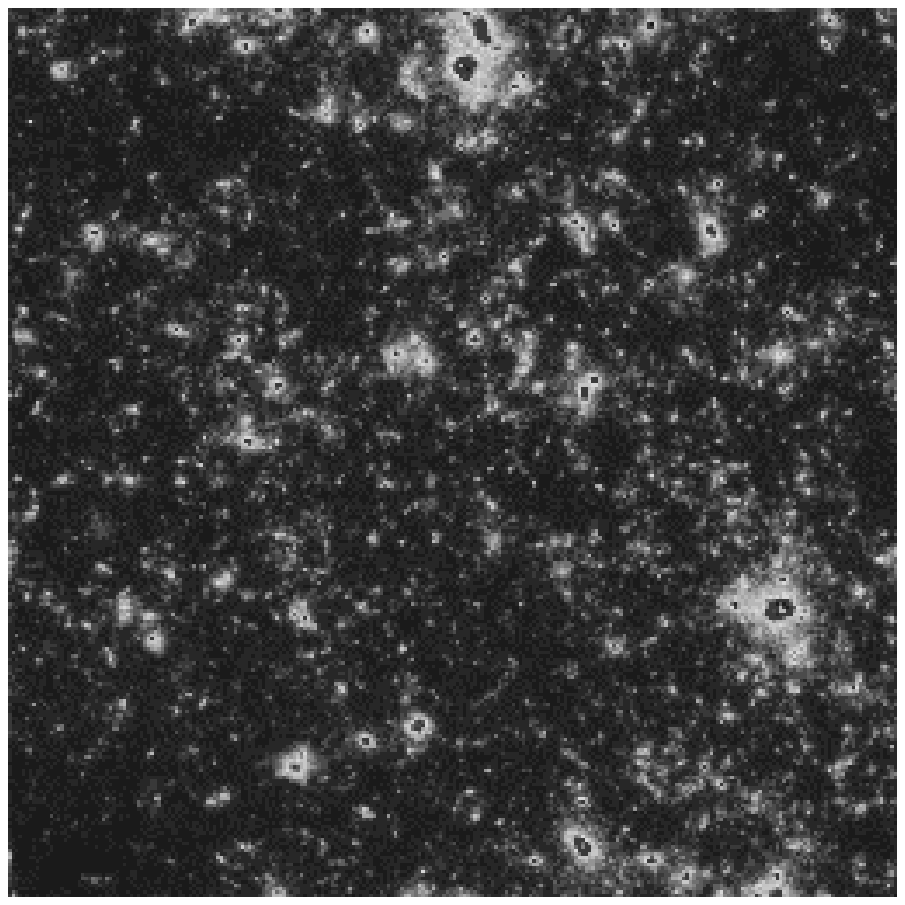}
\label{figmag}
\end{figure}

\begin{figure}[p]
\vspace*{17.5cm}
\caption{The shear pattern on a 1$^\circ$ field for 
Einstein-de Sitter (upper panel) and open (lower panel) models,
corresponding to the magnification patterns of figure \ref{figmag}. 
The open model produces a shear pattern dominated more strongly
by clusters and groups of galaxies. 
}
\includegraphics{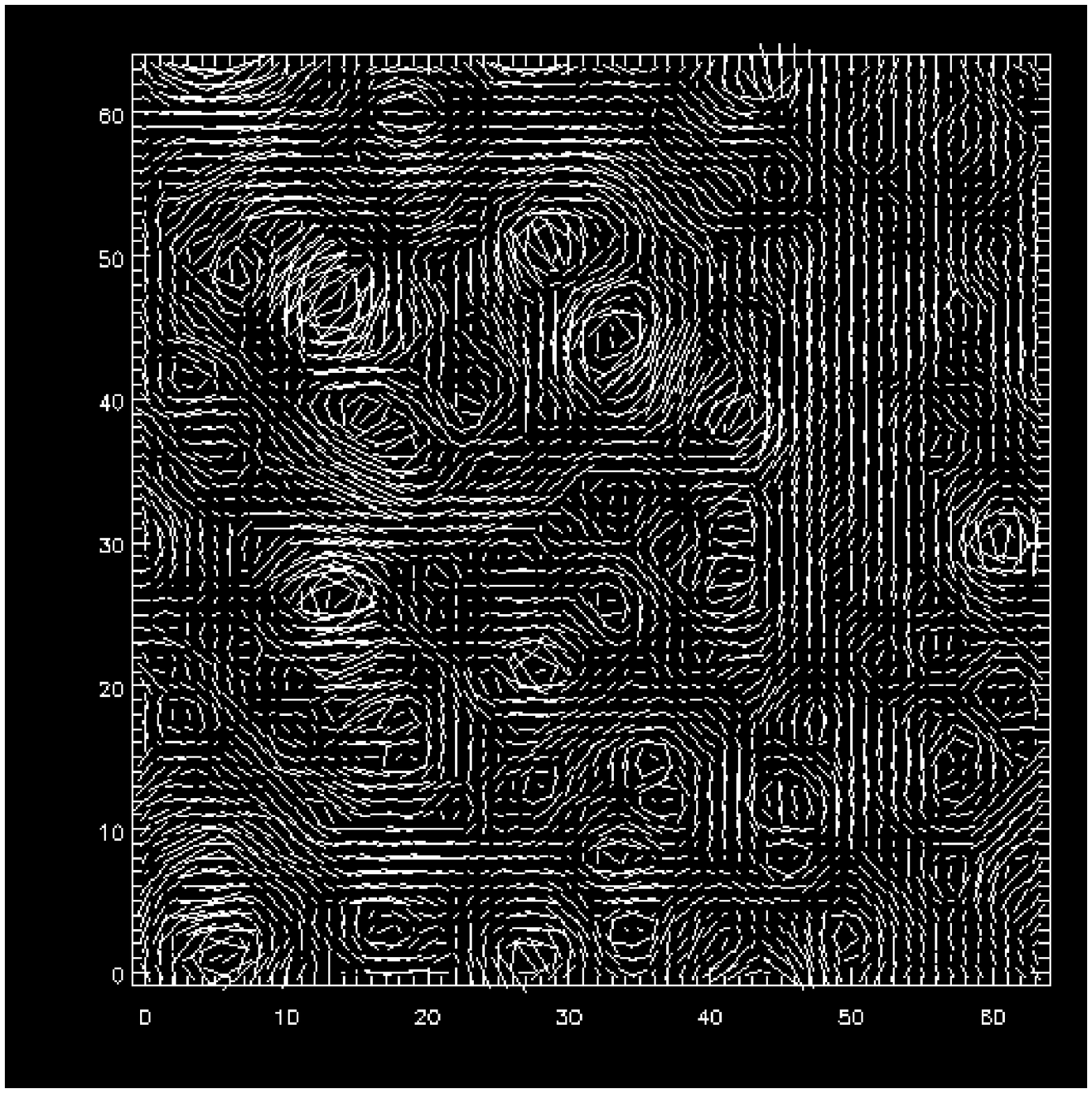}
\includegraphics{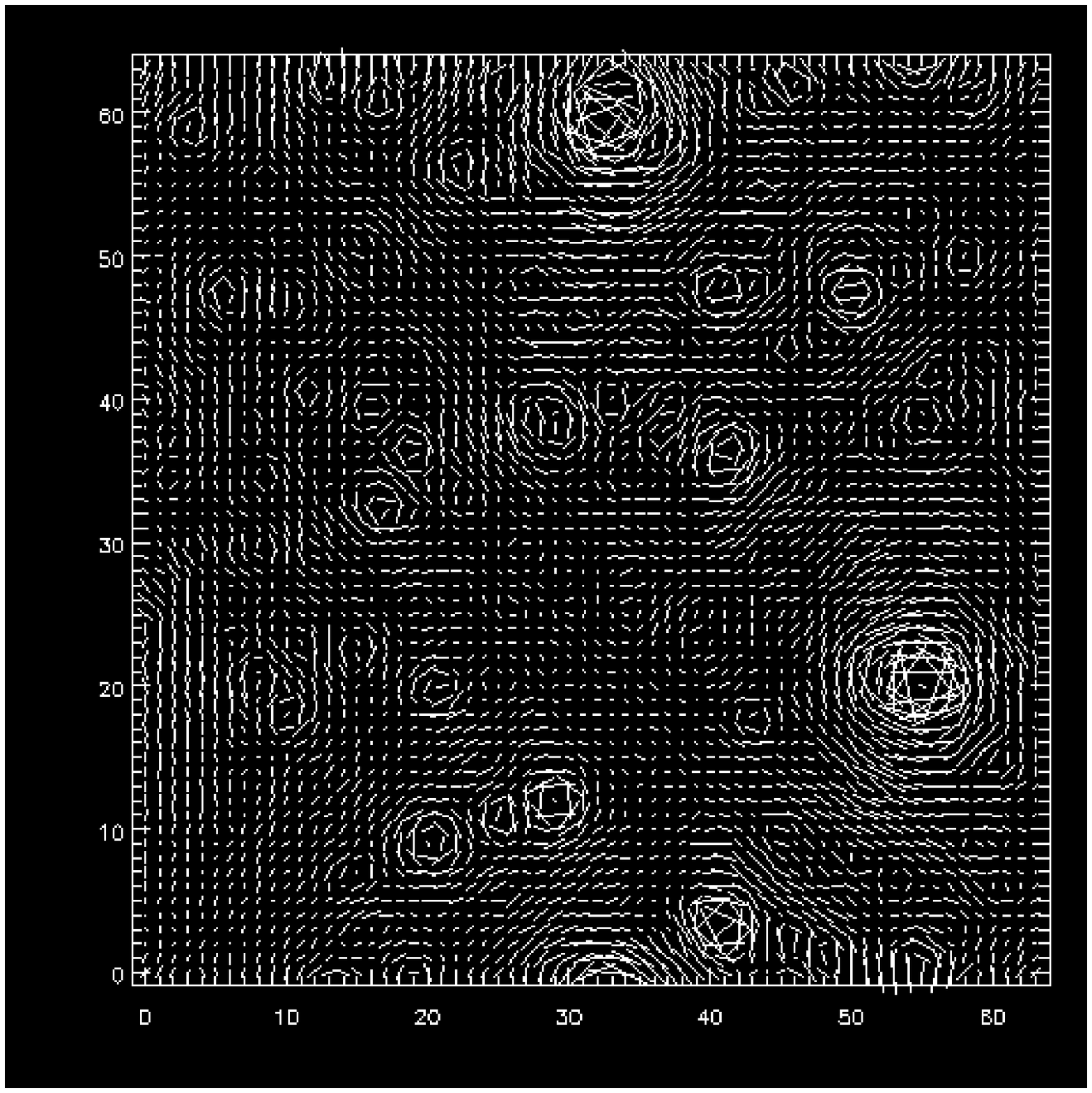}
\label{figshear}
\end{figure}

Maps of the magnification and shear on the source plane are
shown in figures \ref{figmag} and \ref{figshear}. These show
square fields $\sim 1^\circ$ on a side for $z_{\rm source}=1$. 
The qualitative
difference between the $\Omega_{\rm m}=1$ and open models is 
evident in the relative dominance of clusters and groups of galaxies
in the shear field of the open model. The filaments and
other irregular structures visible in the $\Omega_{\rm m}=1$ model are
suppressed in the open models. This occurs because clustering freezes 
in at higher redshift in an open universe. Therefore by $z\simeq 0.3-0.5$, 
the redshifts that dominate the lensing, cluster sized halos have swept 
out most of the matter. The Einstein-de Sitter model on the other
hand has lower normalization, and evolves more rapidly, so that at
$z\simeq 0.3-0.5$ a larger fraction of matter is outside of large
virialized halos. These qualitative differences are reflected
in statistical measures of the shear and magnification. 
Note that throughout the paper we use source galaxies at a single 
redshift. 
We have compared the amplitude of the shear measured for source galaxies 
at $z=1$ with the amplitude measured from a realistic distribution of 
galaxies with the same mean redshift. The difference is small, with
the amplitude in the latter case smaller by less than 10\% for a
distribution given by $n(z)\propto z^2\ exp[-(z/z_0)^{2.5})]$ which 
closely approximates the one estimated by Mobasher et al. (1996) from
the Hubble deep field. 

\begin{figure}[t!]
\vspace*{11cm}
\caption{The power spectra of the convergence $\kappa$, shear $\gamma$ 
and the anti-symmetric part of the Jacobian matrix $\Phi_{12}-\Phi_{21}$
are shown by the solid, dashed and dotted curves respectively. 
The power spectra of $\kappa$ and $\gamma$ are analytically predicted
to be identical. The power spectrum of $\Phi_{12}-\Phi_{21}$
is the same as the power spectrum of the rotational component. 
This is at least 3 orders of magnitude smaller than that of $\kappa$, 
thus validating one aspect of the weak lensing approximation. 
The model used is SCDM, for which nonlinear effects are strongest. 
}
\includegraphics{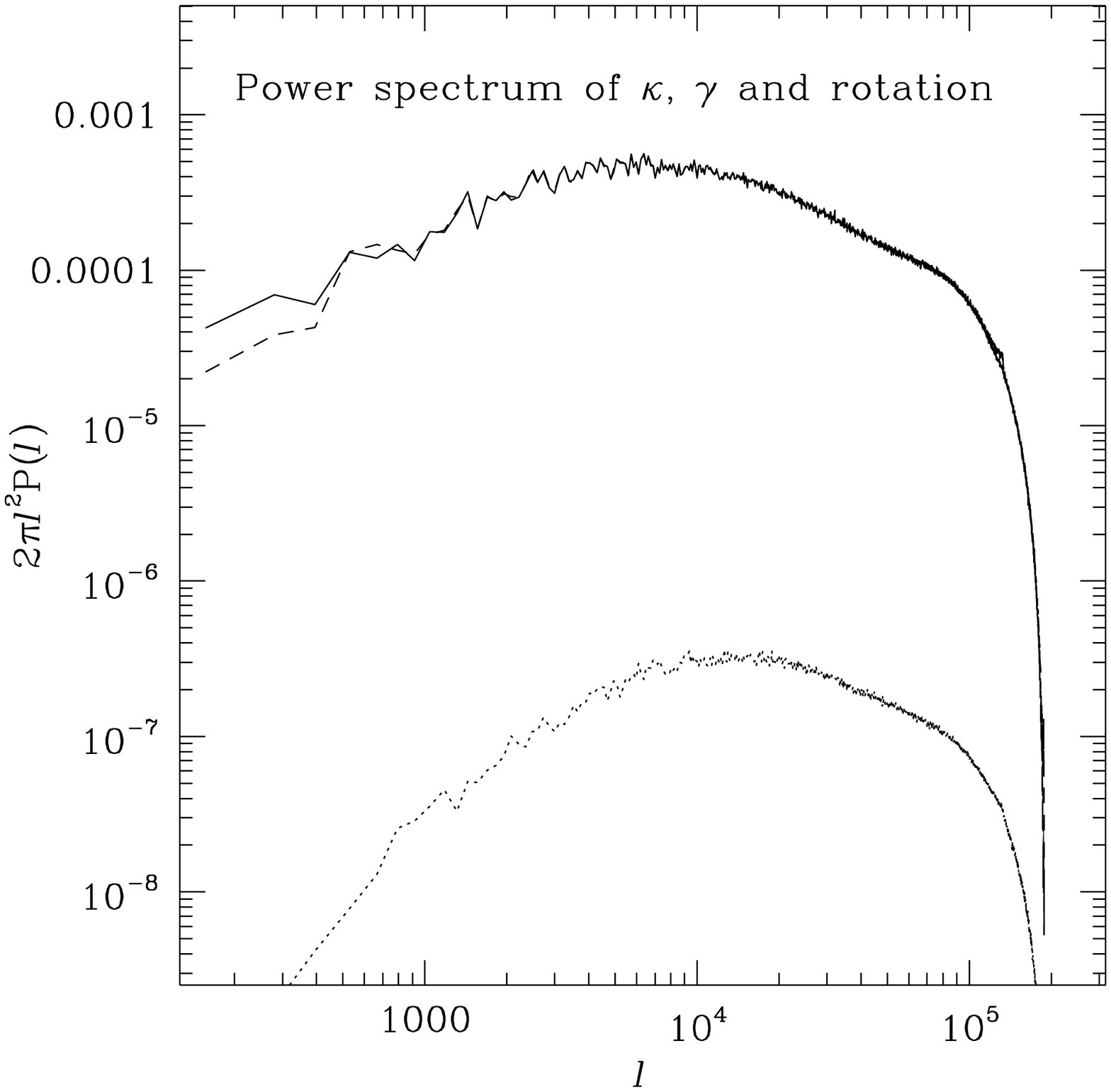}
\label{figantisym}
\end{figure}

As a first check on the accuracy of our results, 
the power spectra of the convergence $\kappa$,  
the shear $\gamma$, and
of the antisymmetric part of $\Phi$ are shown for the SCDM model 
in figure \ref{figantisym}. We have used the SCDM model as it has
more power on small scales and therefore stronger nonlinear effects 
than the other models. 
In the weak lensing approximation 
the power spectra of the first $\kappa$ and $\gamma$ should
be the same -- this is verified by the 
measured spectra in figure \ref{figantisym}, which are nearly identical except
at very small $l$. 
The power spectrum of the anti-symmetric part of the
Jacobian, $(\Phi_{12}-\Phi_{21})$,
on the other hand is smaller by more than 3 orders of magnitude. 
These results imply that 
$\Phi$ can be well approximated as a symmetric
matrix obtained from the 2nd derivatives of a scalar potential. 
This explicitly shows that the corrections produced by quadratic 
terms, such as the one which produces image rotation, are samll.
This numerically verifies one of the assumptions of weak lensing.

\begin{figure}[p!]
\vspace*{17cm}
\caption{The convergence power spectrum for 
different models. For the cosmological model indicated by $\Omega_{\rm m}$ and
$\Gamma$ in the panel, the power spectrum from ray tracing is
compared with the linear (long-dashed) and nonlinear analytical 
(short-dashed) predictions. The power spectra are averaged over
5-10 ray tracing realizations; the error bars show the 1-$\sigma$ 
deviation about the mean. The angular size of the field is
2.8$^\circ$ for the $\Omega_{\rm m}=1$ models, 3.9$^\circ$ for the open
model, and 3.5$^\circ$ for the $\Lambda-$model. 
The upper x-axis labels in the upper panels give the angle in
arcminutes corresponding to $\pi/l$.
}
\includegraphics{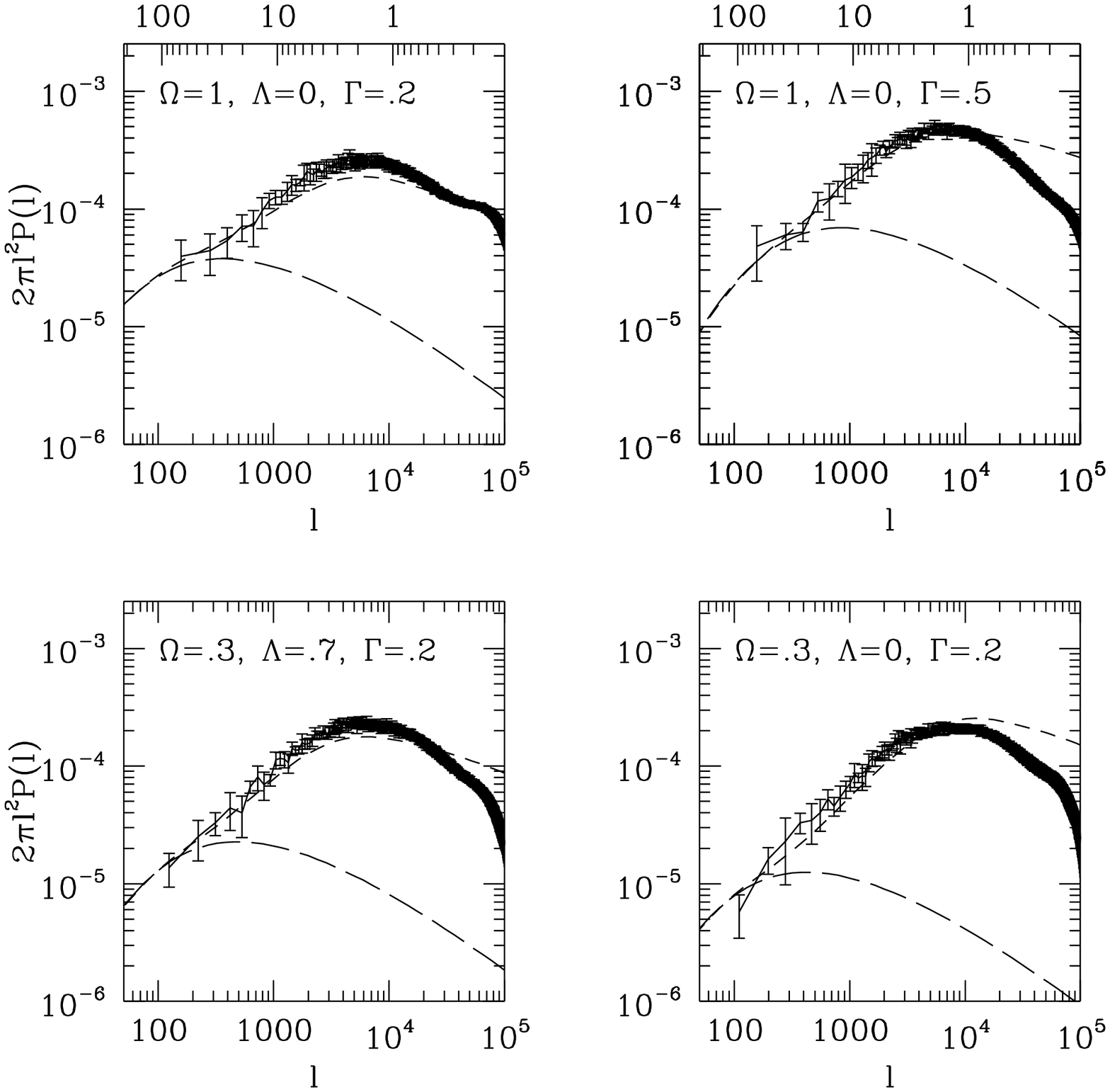}
\label{figpower}
\end{figure}

The convergence power spectrum for the 4 CDM models is compared
to analytical predictions in figure \ref{figpower}.
The error bars show the 1-$\sigma$ dispersion about the mean 
obtained from 5-10 realizations of the ray tracing. 
The dashed curves show analytical
predictions based on the linear and nonlinear power spectrum respectively. 
Details of the nonlinear calculation are given in Jain \& Seljak (1997). 
The x-axis shows the wavenumber in inverse radians. The upper
panels also show the corresponding angle given by $\pi/l$, expressed 
in arcminutes. 
The most striking aspect of figure \ref{figpower} is that the power
spectrum is almost entirely in the nonlinear regime. The measured
power is significantly enhanced over the linear spectrum on all
scales probed by fields of order 3-4$^\circ$ on a side; for $l\sim 10^4$
the enhancement is more than an order of magnitude. Thus with
data from upcoming weak lensing surveys we expect to probe
primarily the nonlinear regime of gravitational 
clustering. 

The agreement of the spectrum measured by ray tracing with the analytical 
predictions of Jain \& Seljak (1997) is very good. There are
slight differences in shape for some of the models but these are
not much larger than the expected accuracy of the analytical 
predictions (Jain, Mo \& White 1995; Peacock \& Dodds 1996). 
Indeed, the 3-dimensional density power spectra of the same 
set of simulations show the same level of
discrepancy between the N-body and the analytical predictions, especially
for the open model (Jenkins et al. 1997). Thus the fitting formulae 
are not sufficiently 
accurate for all models on the smallest scales.  
The discrepancy at high-$l$ is in part due to limited 
small-scale resolution as discussed in the previous section. The
numerical suppression is caused by the finite grid spacing, and 
the smoothing of the projected
density on low-redshift planes (necessary to suppress the white
noise contribution from small particle number). This reduces
the power for $l$ smaller than 1/2 the Nyquist
frequency $\sim 10^5$ radians$^{-1}$, which should set the resolution 
scale from grid effects. 
Aside from the slight
discrepancies in the shape of the power spectra and at high-$l$, the agreement 
with analytical predictions
provides a powerful check on the validity of the weak lensing
assumptions made in the analytical calculation as well as an estimate
of the dynamic range of the ray tracing results. 

\begin{figure}[t!]
\vspace{8cm}
\caption{Comparison of the convergence power spectrum from ray tracing through
PM and P$^3$M simulations. The solid line is for the PM spectrum
and the dashed line for the P$^3$M spectrum. 
} 
\includegraphics{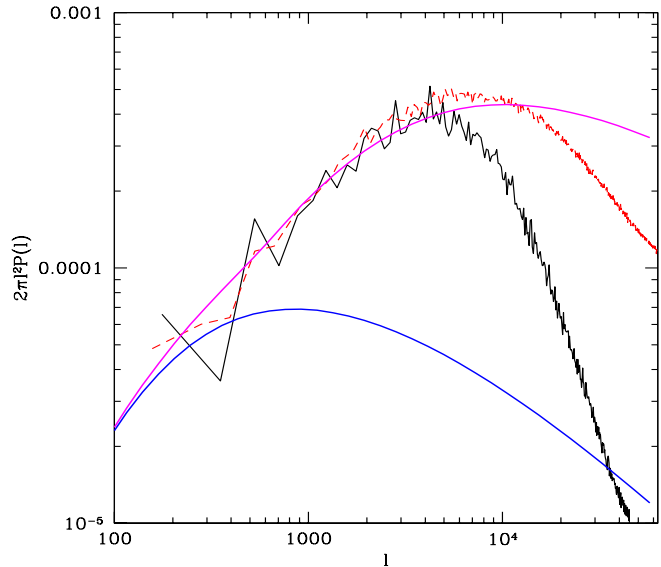}
\label{figpmps}
\end{figure}

Figure \ref{figpmps} shows a comparison of the convergence power spectrum 
from PM and P$^3$M simulations with the analytical predictions. 
Such a comparison is useful for assessing the resolution 
provided by the PM simulations. The simulated spectra agree with
the analytical predictions over two decades in 
wavenumber. The PM spectra suffer from
numerical smoothing on scales about 4 times as large as the P$^3$M. 
This shows that for power spectrum measurements PM simulations 
suffice for $l<5000$. This does not mean that the PM simulations are 
accurate for higher order 
statistics on the same scales: these comparisons are shown in the
following sections.

\subsection{Effects of sample size and noise}

\begin{figure}[p!]
\vspace*{14cm}
\caption{The convergence power spectrum corrected for the mean 
value in a given realization. The non-zero mean value of $\kappa$
within a field $0.4^\circ$ on a side is used to correct the 
computed power spectrum. The correction is made by
adjusting the measured $\kappa$ for the change in the growth 
factor required to produce the finite mean value (see text for details). 
The power spectra shown in
all four panels are for the $\Omega_{\rm m}=0.3$ open CDM model. The upper left
panel shows the uncorrected spectrum, obtained by averaging 64 realizations. 
The lower left panel shows the corrected spectrum for which
the error bars are smaller by nearly a factor of 2 at high $l$. 
The upper right and lower right panels show the power spectra, 
corrected using wrong values of $\Omega_{\rm m}$, 
$\Omega_{\rm m}=1$ and $\Omega_{\rm m}=0.15$ respectively. 
This increases the scatter in the power spectra compared to the 
correct value of $\Omega_{\rm m}$ and could in principle be used
to determine $\Omega_{\rm m}$. 
}
\includegraphics{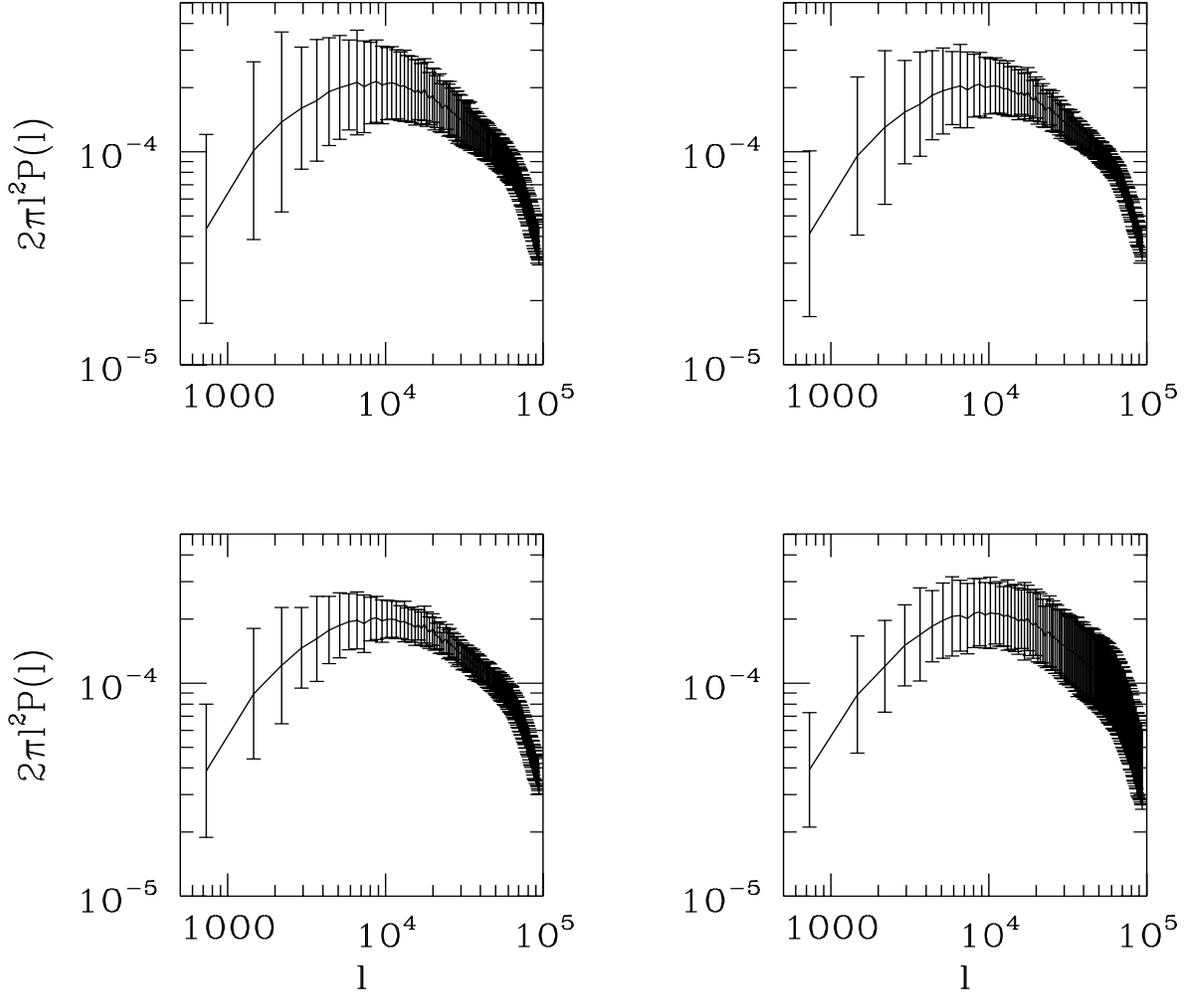}
\label{figcorrected}
\end{figure}

The error bars shown in figure \ref{figpower} indicate the 
dispersion in the measured power for fields of size $\sim 3-4^\circ$
on a side. On these scales the largest modes are linear and 
the fluctuations on large scales are due to sample variance, 
i.e. the small number of modes available to measure the power.
For smaller fields even the largest modes are nonlinear and 
become coupled to the modes on smaller scales. In this case 
the power spectrum variation is no longer dominated
by the finite number of modes, rather it is determined by the strength
of mode coupling which depends on the power on scales of the survey. 
The error bars in the upper left panel of figure \ref{figcorrected}
can be compared with those in figure \ref{figpower} to see the effect
of reducing the field size. 

\begin{figure}[t!]
\vspace*{9cm}
\caption{The convergence power spectrum with noise
due to the intrinsic ellipticities of source galaxies. 
The power spectrum is computed including the intrinsic
ellipticity dispersion on fields with
side-lengths $3.9^\circ$ for the $\Omega_{\rm m}=0.3$ open CDM model.
The error bars are computed by averaging over 10 realizations. 
}
\includegraphics{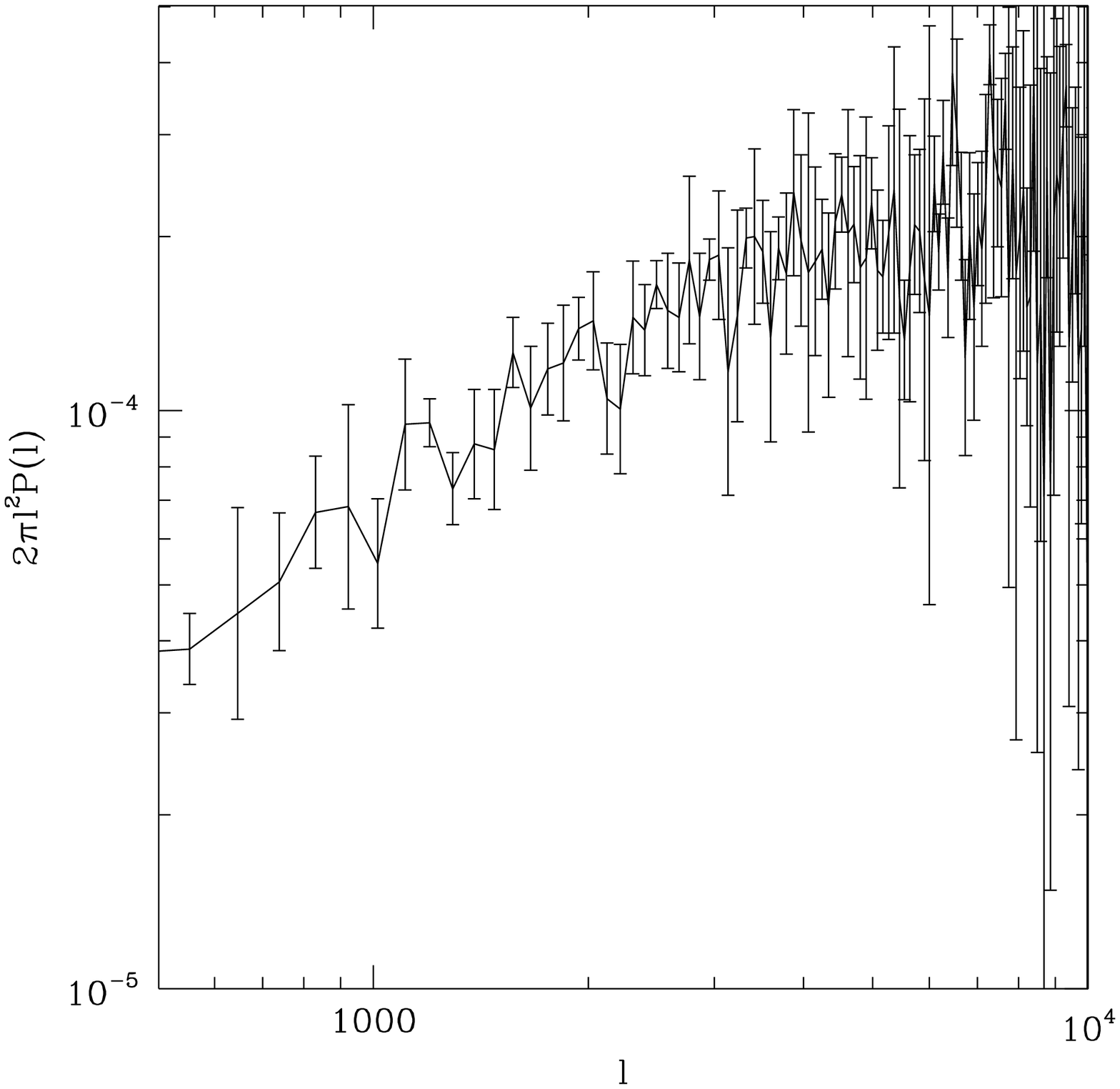}
\label{fignoise}
\end{figure}

\begin{figure}[t!]
\vspace*{9cm}
\caption{The smoothed rms value of $\kappa$ and $\gamma$ versus 
angle $\theta$. The analytical predictions are given by the dashed
curves as in the previous figure. The circles and triangles give
the ray tracing results for one realization, while the solid line
gives the average over four different realizations. There
are negligible differences between realizations in the rms. On large
scales, for diameter $2\, \theta \gsim L/10$, where $L\sim 3-4^\circ$ is 
the side-length of the field, 
the rms measured from ray tracing is suppressed due to the finite size 
of the field. 
} 
\includegraphics{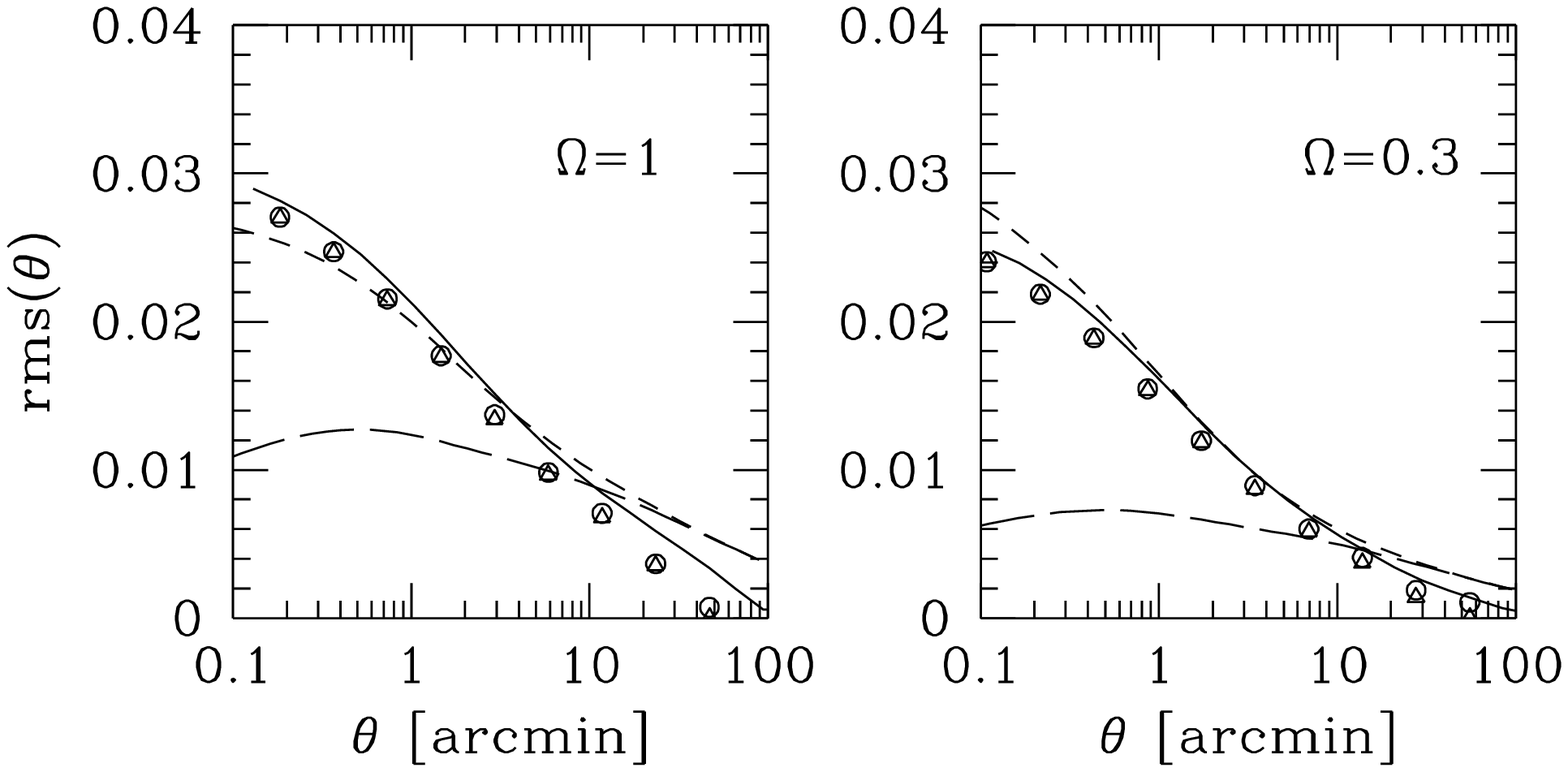}
\label{figrms}
\end{figure}

In figure \ref{figcorrected} we 
explore a way of reducing the scatter
in the measured power on a set of fields of $0.4^\circ$ on a side. The main 
source of scatter
with small field sizes arises from the fact that in a given redshift
bin, the mass density deflecting the light rays is not equal to the global 
mean at that redshift. The effect of the departure from the global 
mean density can be approximated 
by changing the growth factor for the density perturbations within that 
region, i.e. using the growth factor for a different value of
$\Omega_{\rm m}$. This is another way of describing the dominant effect
of nonlinear mode coupling. The mean $\kappa$ measured on the source plane 
is a weighted integral of the density over redshift. It 
therefore provides a measure of the change in the 
density growth factor averaged over redshift with a known window function. 
We could use the value of the measured mean $\kappa$ in that region (which 
could be obtained, for example, from a larger sample, or from the number
counts) to ``correct'' 
the measured power spectrum of $\kappa$ for the fluctuations in
the mean density along the line of sight.
To implement this procedure with observational data would
require assuming a value of $\Omega_{\rm m}$; one could 
repeat the procedure with varying $\Omega_{\rm m}$ until the 
error bars are minimized. In
principle this offers a method of measuring $\Omega_{\rm m}$ from the
fluctuations in the mean $\kappa$ in different fields, but is likely
to give a poorer constraint than using higher moments of the distribution.  
Nevertheless, the fact that for the correct value of $\Omega_{\rm m}$ the 
fluctuations are reduced by nearly a factor of 2 implies that the 
main source of error in the power spectrum are the fluctuations on 
scales larger than the beam, which in this case are of order unity. 
This source of error dominates the effect of finite number of modes,
especially on scales much smaller than the field size for which there
are many modes per wavenumber bin. 

So far we have discussed the effects of sample variance that would
be present even if the data had no noise and the galaxies were intrinsically
circular. The dominant source of noise in observational data are
the intrinsic ellipticities of galaxies. Figure \ref{fignoise}
shows the power spectrum measured from simulated data with 
$2\times 10^5$ galaxies per
square degree whose intrinsic ellipticities are Gaussian distributed
with an rms of 0.4 for each component. Clearly at 
high $l$ the ellipticity noise dominates and prevents the
determination of the 
true power spectrum for $l \gsim 10^4$. For $300\lsim l \lsim 3000$, 
the power spectrum can be measured with an accuracy of a few tens of
percent from of order 10 fields, each a degree on a side. This
wavenumber range corresponds to length scales of $1-10\, h^{-1}$Mpc
at redshifts of 0.3-0.5 that dominate the lensing contribution. 
Thus the mass power spectrum can be estimated most accurately over
this range of scales. To reach length scales approaching $100\, h^{-1}$Mpc
sparse sampling with sufficient number of degree sized fields is required 
(Kaiser 1998). We emphasise that these conclusions have not taken into
account observational noise such as seeing 
that may degrade the signal, as discussed in Section 1. We have compared
our estimates of the sample size with the linear estimates of
Blandford et al. (1991), Kaiser (1992, 1998), Jain \& Seljak (1997) 
and Hu \& Tegmark (1999). 
We find that the error bars of figure \ref{fignoise} fluctuate about
the linear estimates for $l<3000$, while they are dominated by 
shot noise for higher $l$. The estimation of cosmic variance bears
further examination with simulations that do not need to use the
same box multiple times, as this may be affected by corrrelated structures. 

For some data sets it may only be feasible to measure moments
in real space rather than Fourier space. 
The rms of $\gamma$ and $\kappa$, smoothed with a top-hat window in 
real space, is shown in figure \ref{figrms}. The two rms values
on different smoothing angles 
agree very well as expected analytically. The values also
agree well with the nonlinear analytical predictions. The slight
discrepancies at small scales are consistent with the discrepancies
in the power spectra for the same models. 
On large scales the rms falls below the analytical values due to 
the missing power
from long wave modes. The difference is larger than for the power
spectrum because the rms at a given angle involves an integral over 
all wavenumbers. 
The fluctuations between different realizations
are much smaller for the rms than for the power spectrum, 
again because it is an integral over a broad range of wavenumbers. 

\section{Measures of non-Gaussianity}

Given a patch of the sky with measured ellipticities of background
galaxies we would like to extract cosmological 
information with as little loss as
possible. For a Gaussian distribution (likely to be valid on large scales
for popular models of structure formation) 
one only needs to extract the power spectrum from the data, which
completely determines all the statistical properties of weak lensing. 
There is a well defined procedure  for doing this (\cite{seljak98}), based
on a maximum likelihood (ML) method: one writes the full probability
distribution (likelihood function) 
for the measurements as a multivariate Gaussian, whose
unknown parameters are the power spectrum coefficients as a function
of scale. By finding the maximum of the likelihood function we find
the estimated values of the power spectrum coefficients. The 
method is asymptotically unbiased and minimum variance. One can 
also derive a quadratic estimator from this ML method, which is 
easier to compute and leads to the same ML solution (\cite{seljak98}). 
For scales small compared to the survey size the quadratic estimator 
reduces to a simple Fourier method (Kaiser 1998): the Fourier transform coefficients
of the reconstructed $\kappa$ are squared and added together 
for all the modes contributing to a given power spectrum 
bin. The power spectrum estimates are finally obtained by 
subtracting the noise contribution.

For non-Gaussian distributions, which arise due to nonlinear gravitational 
evolution on small scales, the problem of how to optimally extract the
information in the data is more difficult to solve. 
Nonlinear evolution develops correlations between 
Fourier modes which were uncorrelated in the linear regime. This 
mode-mode coupling does not show up in the two point correlator of
Fourier modes because of translational invariance, but is present 
in all the higher moments. The full likelihood function would have 
to describe all these correlations and is therefore not amenable to
analytic expressions.  

Given that the full likelihood function is not achievable,
what is the next best thing to do? In previous studies on nonlinear
clustering various statistical descriptions of non-Gaussianity have 
been developed. 
Among these are moments, N-point correlation functions, the bispectrum, 
Edgeworth expansion of the pdf etc. In 
light of the many possible statistics one can devise, it is difficult 
to take a rigorous, systematic approach. Nevertheless, the fact that 
the non-Gaussian signatures have been
produced by gravity allows one to make some general statements
regarding the merits of different estimators. 
We should emphasize that we 
are interested in the best possible statistic to determine $\Omega_{\rm m}$,
the principal free parameter in addition to the power spectrum that
weak lensing can probe.
In some applications of non-Gaussianity, such as in galaxy clustering, one is 
interested both in the biasing relation and $\Omega_{\rm m}$. To break 
the degeneracy  between the two the data have to be compressed in 
more than a single number (e.g. bispectrum, cumulant correlators
or three-point correlation function; Scoccimarro et al. 1998, Szapudi 1998).
These complications are not present in the case of weak lensing, so 
we can concentrate on the simplest statistics and compress all
the information on $\Omega_{\rm m}$ into a single number.

The first question is whether
one should look for non-Gaussian signatures in Fourier space or in 
real space. We may for example compare the third moment in real space
$\langle \kappa(\bi{r})^3 \rangle$ (or skewness 
$S_3= \langle \kappa(\bi{r})^3 \rangle/\langle 
\kappa(\bi{r})^2\rangle^2$, where the insertion of 
the extra powers of second moment makes $S_3$ independent of 
power spectrum amplitude in perturbation theory)
to the bispectrum, defined in Fourier space as
$\langle {\tilde{\kappa}}(\bi{l}_1){\tilde{\kappa}}(\bi{l}_2)
{\tilde{\kappa}}(\bi{l}_3) \rangle$ with $\bi{l}_1+\bi{l}_2+\bi{l}_3=0$.
We found the bispectrum to be a very noisy statistic, so that even with 
a large number of realizations the signal remained very weak. In 
contrast, the skewness shows a clear signature of nonlinear evolution
and can be measured robustly, as
shown in the next section. This is not surprising, since the bispectrum 
has one more parameter (the shape of the triangle of Fourier modes) and one
has to compress the data from different triangle 
shapes first to obtain more robust 
information. The question of how to combine this information is however
not trivial. The skewness in real space 
is one way to compress this information into a 
single number at a given scale, and while it may not be optimal it has
the advantage of being physical and easy to compute. Note that to 
observe it we need to reconstruct convergence from the shear, which is 
only feasible with sufficiently large fields. As we show below large fields
are required anyways for the signal to be observable, so this is less
of a constraint than it would appear at first. All the results we show
are based on convergence as reconstructed from shear and include any 
additional systematic effects that could in principle arise from this 
procedure. One example of such effects is forcing periodic boundary conditions 
to the data. This generates unphysical structure on the edges of the field,
but the effects are very small and do not show up as a significant effect
in any of the statistical tests we applied. 

The other reason for using real space methods 
is that the action of gravity is quite localized
in real space: it approximately preserves the relative rankings
of density peaks, but enhances their contrast in the distribution
because of nonlinear evolution. This is the reasoning behind the 
Gaussianization procedure (\cite{weinberg}), which attempts to reconstruct 
the primordial density field by mapping the density pdf into a Gaussian. 
Should this mapping 
work perfectly, we would have the full distribution function for 
the data, which would be written as a multivariate Gaussian on a 
nonlinear transformation of the density field (exactly the 
transformation that brings the pdf into a Gaussian form). As discussed
below, we found that this mapping does not work sufficiently well in practice 
to reconstruct the linear distribution. 

Before we proceed, we need to address the issue of
how to combine information obtained from different scales. 
For Gaussian theories information from different scales
can be combined for higher statistical 
significance of the result. This can be done because
in the linear regime modes are independent, allowing information from different
scales to be combined. Once the second moments (or 
power spectrum) are determined there is only one additional parameter
that we aim to determine from weak lensing data, 
the density parameter $\Omega_{\rm m}$ (ignoring the differences
between open and cosmological constant models with the same $\Omega_{\rm m}$).
This parameter can be determined from a range of scales. 
The information from different scales however cannot be combined independently.
If nonlinear evolution were to only enhance the contrast, but not 
change the initial 
ranking order of density values (as assumed in Gaussianization),
then one could argue that the smallest 
scale includes all of the information present in the largest scales as well.

We have tested this by mapping the pdf into a Gaussian on a small scale
(Gaussianize the data)
and then constructing the pdf on a larger smoothing scale. We found that the 
pdf on the larger scale was very close to a Gaussian, so that most of 
the information on the non-Gaussianity of the
pdf was used up by the mapping on the smaller scale.
However, there are still mode-mode
correlations present in the transformed data, so Gaussianization
as a way to obtain their full likelihood function does not work sufficiently
well. These correlations are not accessible by using
the pdf in real space; it would require a much more complicated method 
to add this information to the one from pdf. 

In the absence of noise, the smallest scales are the most 
sensitive to nonlinear effects.
Noise masks information more on small scales, so there exists an optimal scale
where noise is still not dominant and non-Gaussian effects are large 
enough to look for the non-Gaussian signal. The question of how this
optimal scale can be determined is addressed below. The error on the 
estimate of $\Omega_{\rm m}$ is then 
given by the scatter at this optimal smoothing scale. Combining the
data from different smoothing 
scales does not improve the accuracy on $\Omega_{\rm m}$, but 
is still useful as a consistency check.

\section{Results on non-Gaussianity}

\subsection{The one point distribution function (pdf)}

We  now turn to the use of non-Gaussian signatures induced by lensing
to measure cosmological parameters. 
In this section we will discuss the one point distribution function 
(pdf). The pdf has been previously studied by Wambsganss et al. (1995, 
1998) who focussed on the high magnification regime.
We will first present pdf's for various cosmological models
in the absence of noise, highlighting qualitative differences between them.
Simulated data with noise due to the finite number density of galaxies
with large intrinsic ellipticities will then be used to address the question 
of how to extract $\Omega_{\rm m}$ from weak lensing data. 

\begin{figure}[p!]
\vspace*{18 cm}
\caption{Pdf of $\kappa$ at four smoothing scales, 8' (upper left), 4'
(upper right), 2'
(lower left) and
1' (lower right). From top to bottom at the peak values (thin to thick 
lines) are the open model, cosmological constant model, flat model 
(all with $\Gamma=0.21$) and the standard CDM model (with $\Gamma=0.5$).
}
\includegraphics{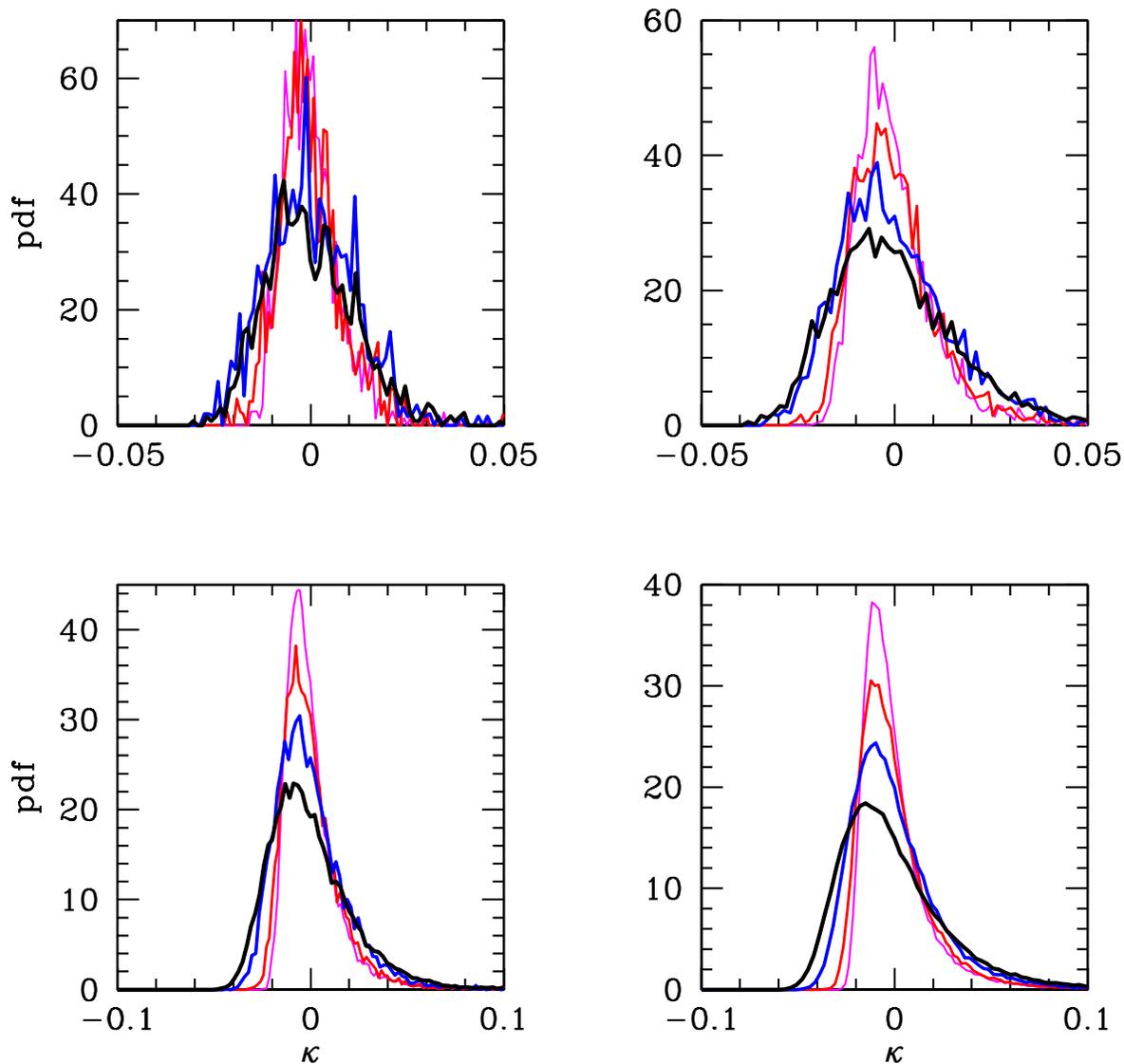}
\label{fig1}
\end{figure}

The qualitative features of the pdf are illustrated in figure \ref{figdeltapdf}
and figure \ref{fig1}. Figure \ref{fig1} shows the pdf for four different 
cosmological models at four different smoothing lengths, assuming the 
sources are at $z=1$. The models are: open $\Omega_{\rm m}=0.3$ model, flat 
$\Omega_{\rm m}=0.3$  model, flat $\Omega_{\rm m}=1.0$ model 
(all with the shape $\Gamma=\Omega_{\rm m}h=
0.21$) and flat $\Omega_{\rm m}=1.0$ with $\Gamma=0.5$ (standard CDM), 
from top to bottom at the peak values of the pdf. 
These are the same models as in figure \ref{figpower} for
the power spectrum. 
At the largest smoothing length (8' top hat radius, upper left) 
the pdf's are close to a Gaussian. There is a 
large scatter because of the small number of cells at this smoothing level
(about 100 per simulation averaged over 10 realizations in this case). 
We use the same smoothing length for all 
the models which will facilitate comparison of the pdf's with noise
added. Since the models are 
normalized to satisfy cluster abundance constraints 
the rms values at a given scale are 
comparable (\cite{JS97}), so that if the pdf's for different models 
were Gaussian, the curves would be very close together. 

As we decrease the smoothing lengths shown in figure \ref{fig1} to 4'
(upper right), 2' (lower left) and 1' (lower right), the pdf becomes more
and more non-Gaussian. This leads to two qualitatively different features.
Overdense regions have
collapsed into dense structures giving rise to a tail of high $\kappa$
that extends to larger and larger values as we reduce the smoothing. At 
the smallest scales $\kappa$ approaches unity in some regions 
where strong lensing can occur. In the quasilinear 
regime on larger scales the amount of non-Gaussianity depends on the rms 
in the density perturbations at the smoothing scale. We 
may loosely describe the smoothing scale in 3-d as the angular 
smoothing scale in 2-d multiplied with the half distance to the galaxies,
where the lensing probability peaks. Comparing between an open and flat model
with the same rms $\kappa$ one sees that a larger rms $\delta$ is needed
in the open model, to compensate for the weaker focussing effect due
to the smaller density $\rho$. 
A larger rms $\delta$ implies non-Gaussian signatures are more important 
and lead to an enhancement of the pdf at large positive $\kappa$. 
This high density tail is also the essence of the moments method
to determine $\Omega_{\rm m}$ (next section; see also 
\cite{Bernardeau97,JS97}). Note that the pdf's for open and flat 
low density models are visually quite similar. This is quantified
in more detail below.

For negative $\kappa$ there is an even more robust
qualitative difference. Nonlinear evolution sweeps matter 
from underdense regions and puts it into collapsed halos or filaments
and sheets connecting them. This happens very early 
in an open universe, so that there are large regions of empty space
throughout the line of sight.
Many lines of sight propagate through empty space in 
such a model and the pdf peaks sharply very close to the smallest possible
value at small smoothing lengths and drops to 0
rapidly below that. In a flat universe clustering evolves more rapidly
at low redshift, so that the universe is more weakly clustered at
$z\lsim 1$ compared to an open universe. 
In addition, cluster abundance normalization gives a lower amplitude
for fluctuations in a flat universe 
at the 10 $h^{-1}$Mpc scale even at the present epoch.
For a flat universe the peak in the pdf is thus not close
to the minimum value and its shape is less steep, at least for smoothing
scales of 1' and larger. 

\begin{figure}[p!]
\vspace*{18 cm}
\caption{Pdf of $\gamma$ smoothed on a scale of 2'. The upper
two panels show the pdf and cumulative pdf of the open (dashed) and 
flat (solid) models with $\Gamma=0.21$ spectra. The lower left panel
shows the difference of the pdf from a Gaussian with the same
rms. The lower right panel shows the ratio with the same Gaussian
for the two models. 
}
\includegraphics{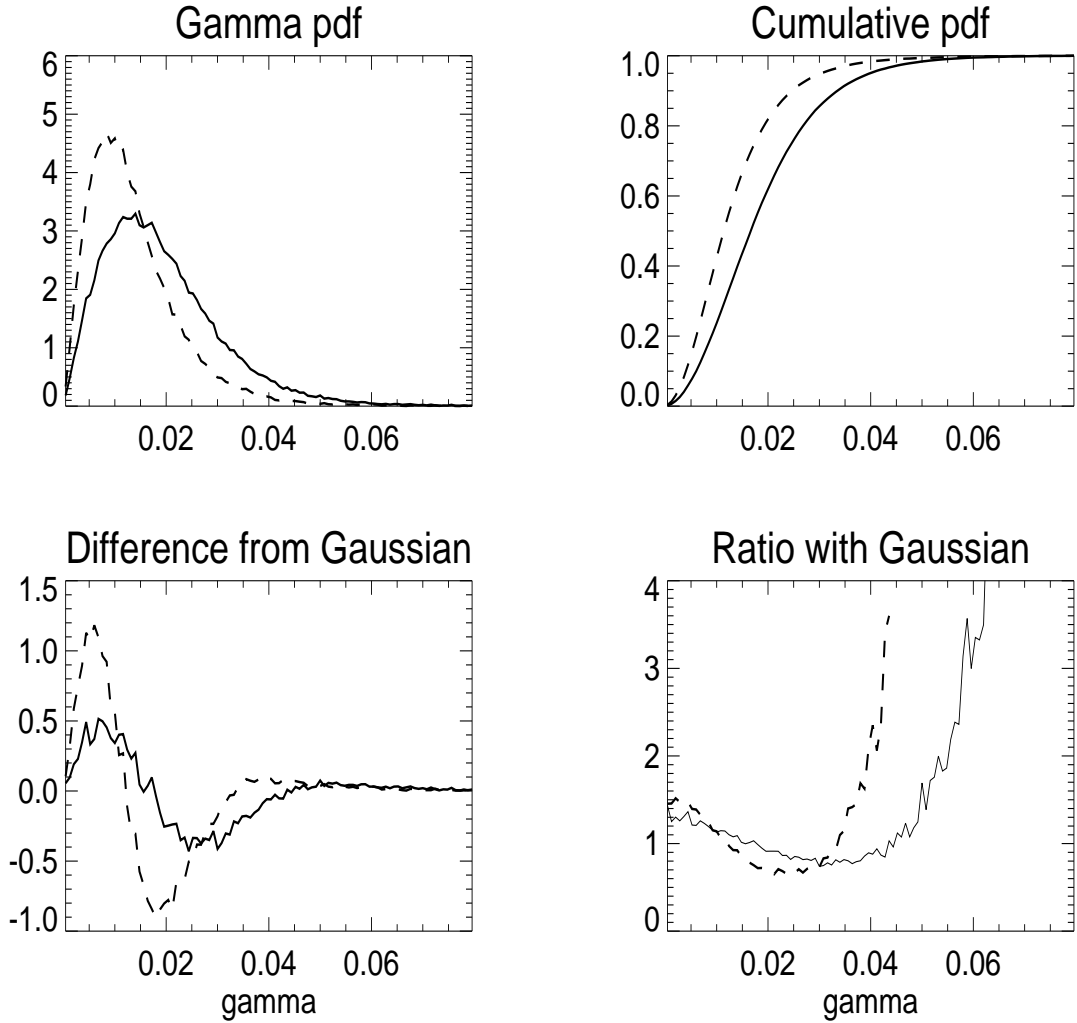}
\label{fig1gamma}
\end{figure}

The minimum value of reconstructed $\kappa$  
is proportional to the density parameter $\Omega_{\rm m}$. 
This is the reason that the minimum value of 
$\kappa$ in an open universe is much smaller than the corresponding
value in a flat universe at small smoothing angles, even if the rms is 
comparable or even larger. The minimum value of $\kappa$ can be used as 
a direct measure of the density in the universe, once the redshift distribution
of background galaxies is known. This does depend somewhat on 
the geometry, as the pathlength and angular scale differ between 
open and cosmological constant models with the same $\Omega_{\rm m}$.
The empty beam gives the minimum value of $\kappa=-0.12$ for $\Omega_{\rm m}=1$
compared to $-0.04$ for the open $\Omega_{\rm m}=0.3$ model
and $-0.06$ for the cosmological constant model.
Note however that while in the open 
universe the measured minimum value is close to the above value
(because there are many beams which are almost empty), 
in the flat universe the lowest measured value is 
far from the empty beam value.
The difference in $\kappa_{{\rm min}}$ 
is therefore smaller and one has to use the full shape of the pdf in 
addition to the minimum value to estimate $\Omega_{\rm m}$. 
\footnote{We found that even
at the highest resolution of our simulation 
of flat universe there are no  
lines of sight with completely empty beams. One cannot  
exclude the possibility that at even smaller scales the universe does
become empty for some or even the majority 
of lines of sight. An independent argument against the majority of 
lines of sight being empty on smalle scales is 
based on analytical rms calculations that include nonlinear evolution of the
power spectrum (\cite{JS97}). The rms is significantly
smaller than the minimum $\kappa$, which would not be the case if most
lines of sight were empty. 
} 

For completeness we show the pdf of $\gamma$ in figure \ref{fig1gamma}. 
This pdf is harder to interpret than that of $\kappa$, but is simpler
to construct from observational data, since it does not require large
enough fields to reconstruct $\kappa$ accurately. It is evident that
even directly with data on $\gamma$ the pdf can discriminate between
models with different values of $\Omega_{\rm m}$. A complete study
of the best way to extract $\Omega_{\rm m}$ from the pdf of $\gamma$
constructed from noisy data will not be presented here. We proceed 
instead with a detailed exploration of the non-Gaussian features in the 
pdf of the reconstructed $\kappa$. 

\begin{figure}[p!]
\vspace*{19 cm}
\caption{Pdf's for the open $\Omega_{\rm m}=0.3$ model for sources at 
four redshifts: from top to bottom at the peak values (thin to 
thick lines) 
$z=0.5$, $z=1$, $z=2$ and $z=3$. The smoothing lengths in the different
panels are the same as in figure \ref{fig1}.
}
\includegraphics{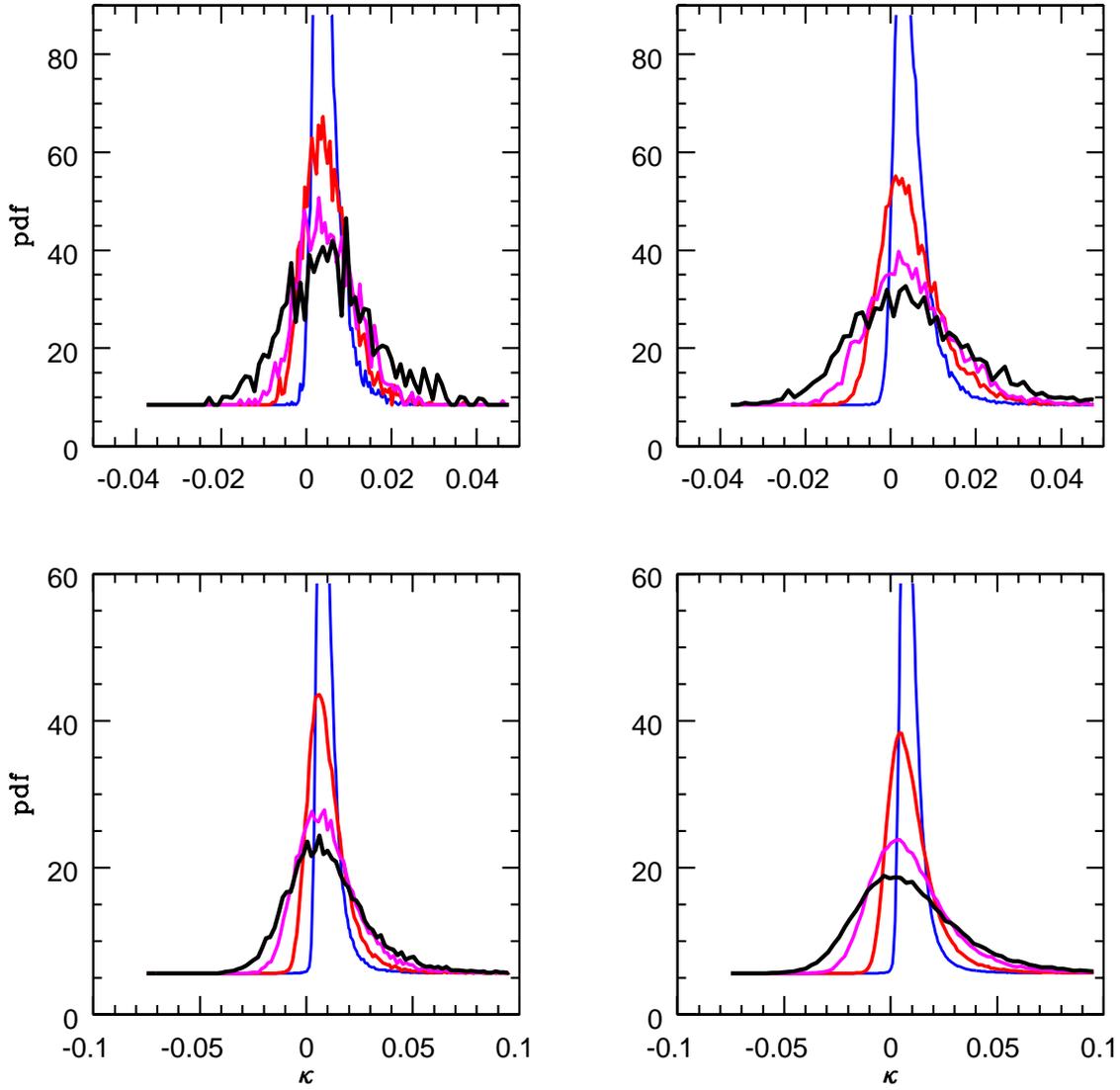}
\label{fig9}
\end{figure}

Figure \ref{fig1} assumes all the galaxies are at $z=1$. Figure 
\ref{fig9} shows how the pdf changes with the redshift of galaxies 
for the open model. From top to bottom are pdf's for sources at 
$z=0.5$, $z=1$, $z=2$ and $z=3$. The deviations from Gaussianity are larger 
at low $z$. This is because as $z$ increases, one projects a larger number
of independent regions. By the central limit theorem 
this leads to a Gaussian distribution, even though the 3-d pdf of 
density is very non-Gaussian (approximately a 
lognormal). On the other hand,
the rms of the convergence (the width of the pdf) 
is increasing with redshift, so that the observability of the
effect is likely to increase with redshift, because noise masks the
non-Gaussian signal when the rms is small.

\subsection{Pdf of $\kappa$ from simulated noisy data}

From the considerations above it follows that 
if one could measure the pdf on small scales directly one could easily 
determine $\Omega_{\rm m}$. Unfortunately, this is not feasible because
of noise. As discussed in the introduction the main source of random noise 
is the intrinsic ellipticities of background galaxies. In addition, 
at faint magnitudes measurement errors can cause a significant error in the
ellipticity determination. We will model the ellipticity noise
as a Gaussian with 0.4 dispersion for each component, roughly in 
agreement with observational constraints. The Gaussian assumption
for the noise is not crucial in this context. The data used is first
smoothed, which makes the noise more Gaussian. In addition, even 
if the noise is non-Gaussian it will not create an asymmetry between positive 
and negative $\kappa$, so at least for statistics like the odd moments 
this will not introduce a bias. In practice one should use 
the actual statistical distribution of ellipticities, which can be 
obtained from the data itself under the assumption that in the 
unsmoothed data, ellipticities induced by weak lensing are negligibly small. 

\begin{figure}[t!]
\vspace*{8 cm}
\caption{The reconstructed convergence field from 
ellipticity data on a single 3$^\circ$ field, smoothed on the
scale where the noise power spectrum exceeds the signal power spectrum. 
The right panel shows the reconstructed field using
randomly distributed galaxies at $z=1$, with each component of
the intrinsic ellipticity randomly drawn from a
Gaussian distribution (details are given in the text). 
The left panel shows the field without noise. 
}
\includegraphics{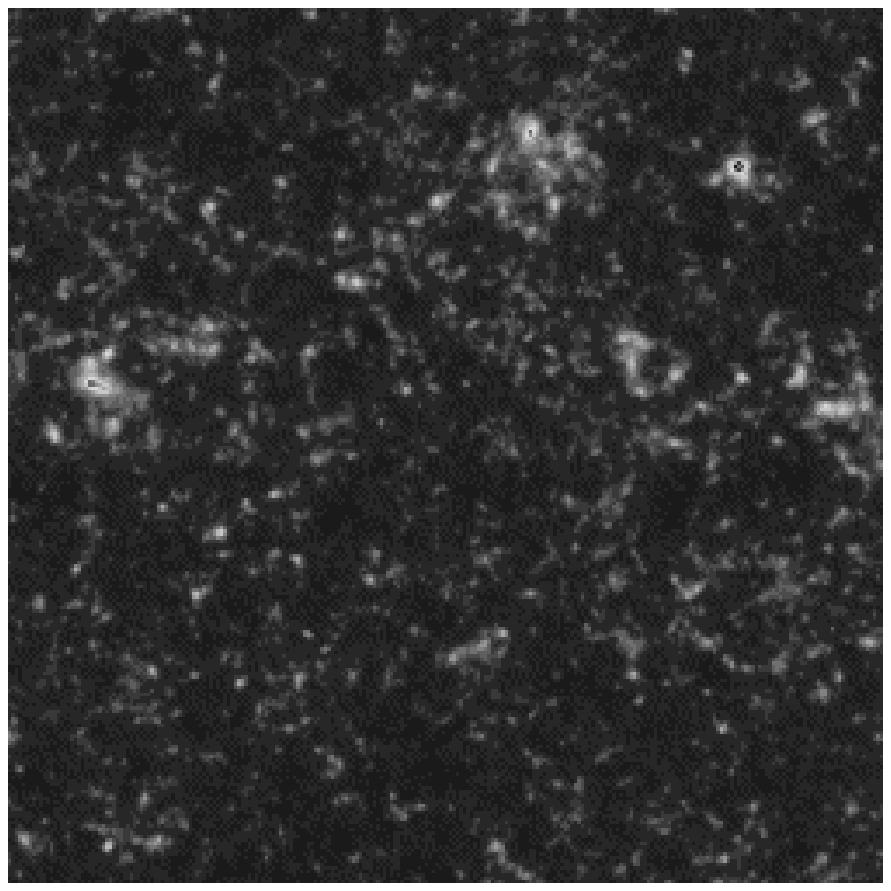}
\includegraphics{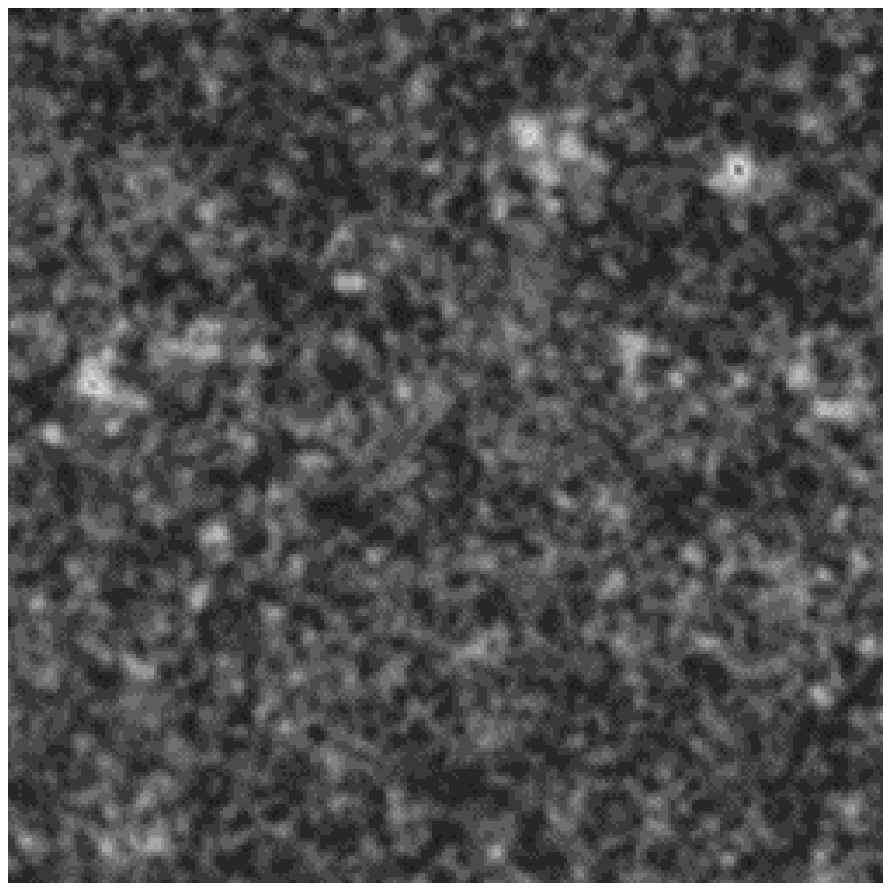}
\label{figkappa}
\end{figure}

The number density of galaxies depends on the limiting magnitude of the 
observation. Here we assume $2\times 10^5$ galaxies per square degree, 
which is at the limit of ground based observations with several hour exposures
(e.g. Luppino \& Kaiser 1997). 
As before all the galaxies are placed at $z=1$. 
Our goal is to check using realistic examples if the signal is detectable 
and what method is best to analyze the data; the details will clearly 
depend on the specifics of the survey one has in mind.  
To generate a noisy version of the 
data we randomly generate galaxy positions on a square grid of
a given size and add the  
weak lensing shear to a random ellipticity
component in each of them. We then reconstruct $\kappa$ from these
noisy data using the
relation between the Fourier transforms of $\kappa$ and $\bi\gamma$,
again assuming periodic boundary conditions:
\begin{equation}
\hat\kappa(\vec l)=\frac{l_1^2-l_2^2}{l^2}\ \hat\gamma_1(\vec l)\ + \ 
\frac{2l_1 l_2}{l^2}\ \hat\gamma_2(\vec l)\ .
\label{kappaft}
\end{equation}
The reconstructed $\kappa$ field smoothed at the scale 
where noise exceeds the signal is compared to the same field
without noise in figure \ref{figkappa}. 

\begin{figure}[p!]
\vspace*{18 cm}
\caption{The pdf of $\kappa$ for different smoothing scales and
cosmological models, as in figure 1, but with noise added. 
}
\includegraphics{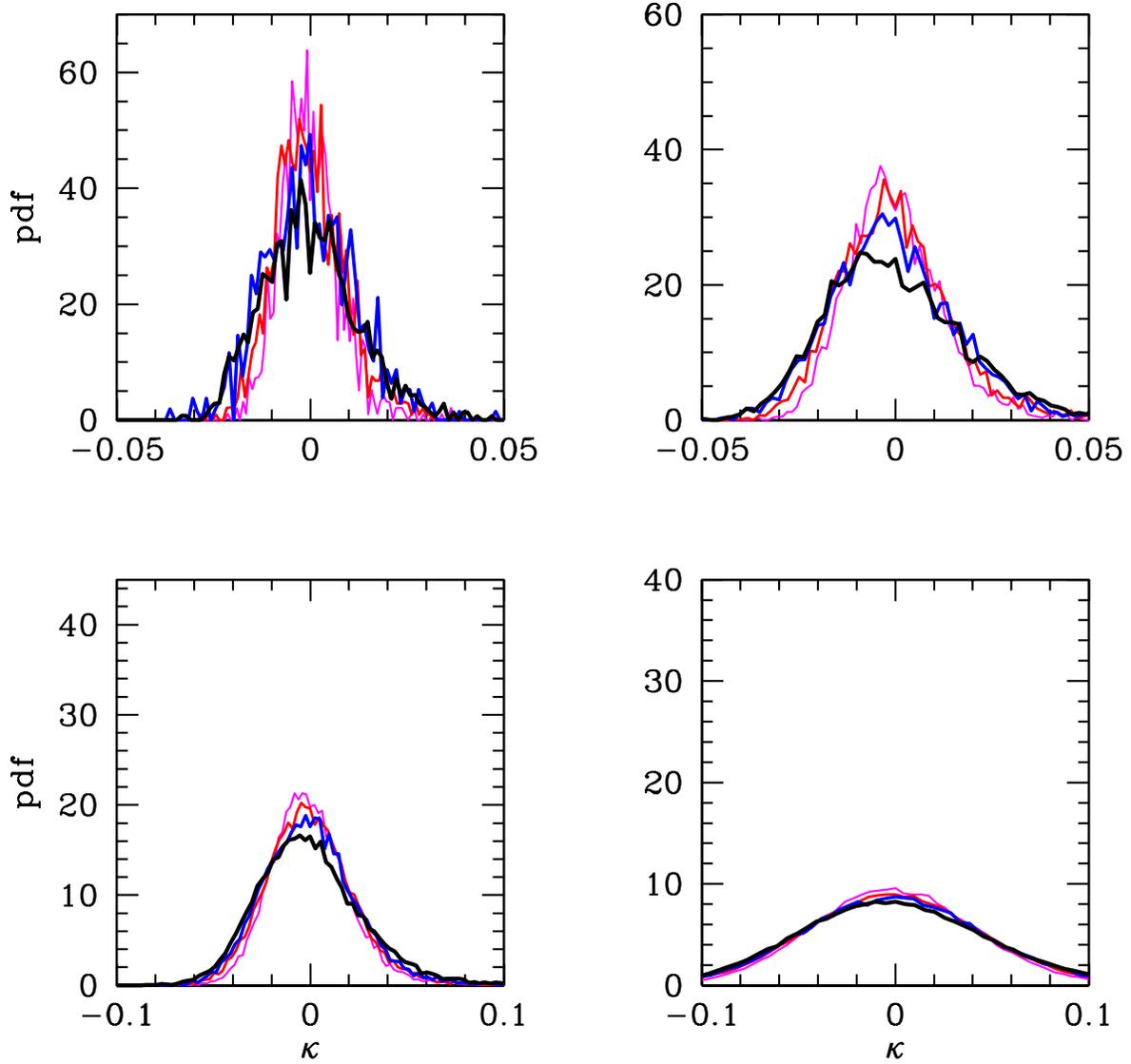}
\label{fig2}
\end{figure}

Figure \ref{fig2} shows pdf's for all the models in figure \ref{fig1},
with noise added. On large 
scales the noise is negligible and the pdf's are similar to the no noise case.
On very small scales the noise is large and dominates over the cosmological
signal, so that the pdf is simply a Gaussian with 
rms$=0.4/N_{\rm gal}^{1/2}$, 
where $N_{\rm gal}$ is the mean number of galaxies in a smoothing patch. 
In both limits the distribution is Gaussian, so
we cannot learn anything about non-Gaussianity. In the intermediate
regime, on scales of $2'$ and $4'$, differences between models are
still visible. One can hope that there is sufficient signal to noise to 
extract useful information from the data. We discuss next what the 
optimal scale is and what statistics are best to extract the signal 
from the data.

\subsection{Moments and other reduced statistics}

Having argued that real space statistics exhibit stronger non-Gaussian
signatures we need to decide which to choose. The first set of statistics 
to try are the moments 
$S_N=\langle \kappa(\bi{r})^N\rangle/\langle \kappa(\bi{r})^2\rangle^{N-1}$,
for $N=3$ this is the skewness already introduced above.
Although one may argue that the pdf contains more information 
than any of the reduced statistics, the moments can nevertheless 
be useful because of their simplicity and relative model independence.
One advantage of moments is the availability of perturbation theory, with
which one can compare the observations on large scales. 
The main disadvantage is that the
moments are sensitive to rare events in the tails of
the distribution. The results can be 
strongly dependent on the presence or absence of a few clusters
(\cite{Colombi94,Szapudi96}). This has two consequences: first, 
sampling variance becomes a more important source of noise
as we progress from lower to higher order moments. Second, numerical
accuracies of simulation codes become an issue and it becomes difficult
to accurately calibrate the observations in the nonlinear regime. 
This has prompted
some workers to try alternatives such as the moments of
absolute values (Nusser \& Dekel 1993; Juszkiewicz et al. 1995). 
These are less sensitive to the rare events, but are more sensitive
to the presence of noise as shown below.

\begin{figure}[p!]
\vspace*{17.5 cm}
\caption{$S_3$ statistics from 4 different measures: second moment over
positive values (upper left), negative values (upper right), 
difference between negative and positive values (lower left) and
the third moment (lower right). From top to bottom (thin to thick 
lines) are the open model, cosmological constant model, flat model 
(all with $\Gamma=0.21$) and the standard CDM model (with $\Gamma=0.5$).
Dashed curves are perturbation theory results. Errors are based on 
the scatter of 5-10 realizations.}
\includegraphics{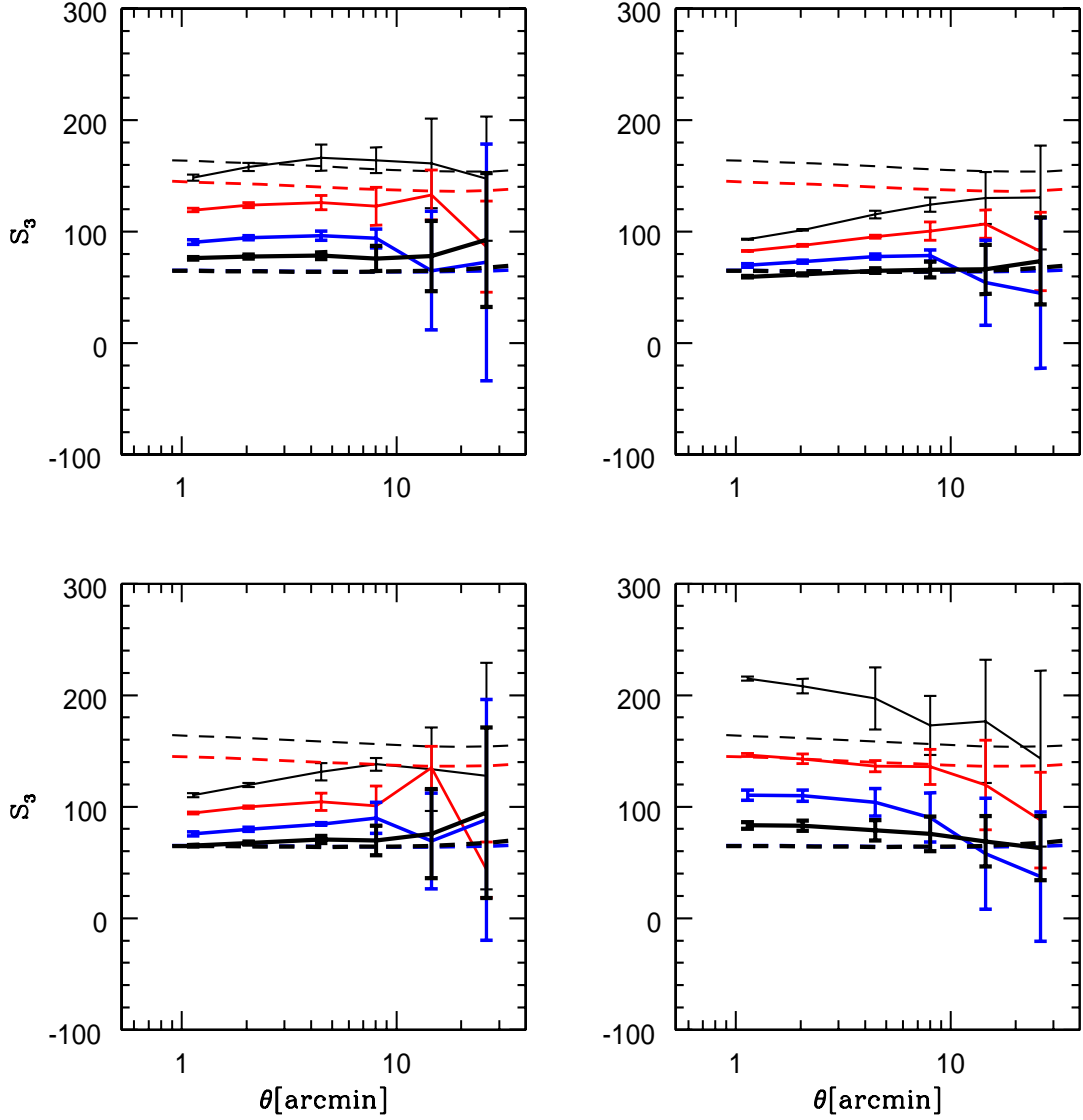}
\label{fig6}
\end{figure}

Based on these arguments 
we will concentrate on four low order moments, all of which can be 
expressed in terms of
the lowest nonvanishing contribution from the third moment $S_3=
\langle \kappa^3 \rangle /\sigma_{\kappa}^4$ and so can be 
compared to perturbation theory predictions. 
The simplest way to compute $S_3$ is simply by its definition above. This 
statistic is plotted in the lower right panel of figure \ref{fig6}. 
Shown are simulation results for the same 
cosmological models as in figure \ref{fig1}, together with the perturbation
theory results shown as the dashed lines. The perturbation 
theory values have been analytically computed using the expressions given
in Bernardeau et al. (1997) and Jain \& Seljak (1997). 
The important point is that $S_3 \propto \Omega_{\rm m}^{-1}F(n)$,
where $F(n)$ is a weak function of the shape of the power spectrum, but 
not of its amplitude. That this function only weakly depends on the 
spectral shape is seen from the comparison of $\Gamma=0.21$ and $\Gamma=0.5$
models in figure \ref{fig6}, 
both of which give almost identical perturbation theory results.
At a given redshift the perturbation theory predictions depend 
somewhat on the assumed geometry and pathlength, 
causing a difference of order 15\% between 
a cosmological constant model and an open model with the same 
matter density $\Omega_{\rm m}$. 

The results of N-body simulations agree with perturbation
theory on large scales, although the scatter in individual realizations 
is large even in the absence of noise. On small scales $S_3$ rises
above its perturbation theory value by different amounts depending 
on the model. For the open model the rise above the perturbation 
theory value is a factor of 2 on very small scales, 
while for the cosmological constant model there is no rise
at all. Moreover, there is also a dependence on the shape of 
the power spectrum, resulting in different predictions for 
$S_3$ between $\Gamma=0.21$ and $\Gamma=0.5$ models. 
Nevertheless, the difference between the low and high 
$\Omega_{\rm m}$ models is clearly seen. Thus in the absence of noise 
due to the finite number of galaxies, the statistical significance 
would be enormous. The interpretation of the result 
however is complicated by the shape and geometry
dependences on very small scales. For this reason it is better to 
use the data on more intermediate scales, where the deviations from 
perturbation theory are smaller. This conclusion will be further 
strengthened when we discuss the effects of noise.

Two other statistics related to $S_3$ in perturbation theory are
$\langle \kappa |\kappa| \rangle_{\kappa>0} $ and 
$\langle \kappa |\kappa| \rangle_{\kappa<0} $. 
These statistics have been suggested by Nusser \& Dekel (1993) and 
Juszkiewicz et al. (1995) as a way to reduce the sampling variance 
from rare events which populate the tails of the distribution and 
strongly affect the higher moments. Because of quadratic 
weighting one can expect these statistics to be more robust.
The contributions from positive and negative $\kappa$ are separated
because they probe different physical regions, halos and voids, respectively.
We use  the Edgeworth expansion of the pdf around the Gaussian to compute 
the perturbation theory values of these statistics (Juszkiewicz et al. 1995). 
At lowest order, the Edgeworth expansion is given by 
\begin{equation}
p(\kappa)= { 1 \over \sqrt{2 \pi}\sigma_{\kappa}} 
e^{-\kappa^2/2\sigma_{\kappa}^2} \left[ 1+ {1 \over 6} S_3\sigma_{\kappa}
H_3(\kappa/\sigma_{\kappa})\right],
\label{edge}
\end{equation}
where $H_3(x)=x^3-3x$ is the 3rd order Hermite polynomial.
The statistic introduced above is then
\begin{equation}
\langle \kappa |\kappa| \rangle_{\kappa>0} =
\int_0^{\infty}p(\kappa)\kappa^2 d\kappa=
\sqrt{2 \over 9\pi} S_3 \sigma_{\kappa}^3 \ . 
\end{equation}
For the $\kappa<0$ statistic one obtains the same perturbation theory value.
In figure \ref{fig6}
we plot $S_3$ obtained using this method separately 
for positive values (upper
left plot) and negative values (upper right).
The predictions of perturbation theory again 
agree with N-body simulations on large scales, albeit with large 
scatter. On small scales 
the positive $\kappa$ statistic 
remains approximately constant, while the negative $\kappa$ statistic
drops in value towards 0. This is expected, because 
$\kappa$ is bounded from below, so the rms over negative
values cannot exceed the minimum value, while the  full rms in the
denominator continues to grow on small scales.

The fourth statistic we measured is $-\langle {\rm sign}(\kappa)\rangle$, 
which is the sum over all negative $\kappa$ minus the sum over all
positive $\kappa$, normalized by the total number of $\kappa$.
This statistic is the least sensitive to the tails and should be 
very robust to the details in high density regions.
Again using the Edgeworth expansion we find at lowest order 
\begin{equation}
-\langle {\rm sign}(\kappa)\rangle=\int_{-\infty}^0p(\kappa)d\kappa-
\int_0^{\infty}p(\kappa)d\kappa=
(18\pi)^{-1/2}S_3\sigma_{\kappa}.
\end{equation}
The $S_3$ measured from this statistic is plotted in the 
lower left of figure \ref{fig1}. 
In the absence of noise it shows the same properties as the 
statistics discussed above. It agrees with perturbation theory on 
large scales, but with large scatter. On small scales it does not differ 
significantly from the perturbation theory value and shows remarkably 
small scatter. However, these appealing properties do not survive once 
noise is added, as discussed below.

\subsection{$S_3$ from simulated noisy data}

\begin{figure}[p!]
\vspace*{19 cm}
\caption{$S_3$ statistics, as in the previous figure, but 
with ellipticity noise added to the simulations.}
\includegraphics{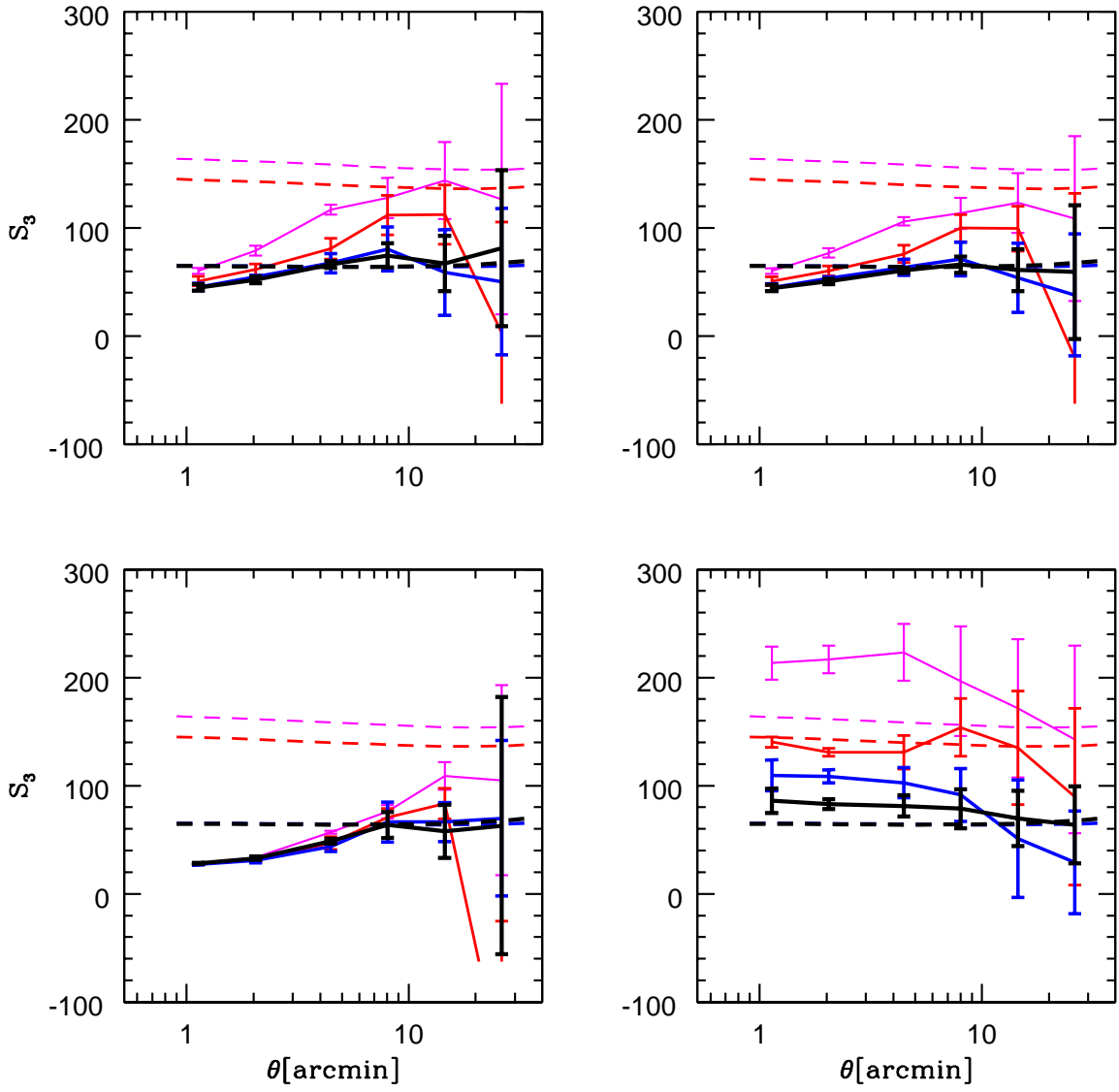}
\label{fig7}
\end{figure}

We now add noise due to the finite number of source galaxies 
to the simulated maps and 
compute the same statistics as in figure \ref{fig6}
in a $3^{\circ} \times 3^{\circ}$ field, shown in figure \ref{fig7}. 
Noise introduces significant scatter and makes 
some of the statistics biased. The third moment 
(lower right panel) has the advantage that it is unbiased
even in the presence of noise. To show this we can write the measured $\kappa=
\kappa_s+n$, where $\kappa_s$ is the true convergence and $n$ is the 
noise component. The former has a probability distribution $p(\kappa_s)$,
while the noise we assume to be Gaussian distributed as
$p_n(n)=(2\pi \sigma_n^2)^{-1/2}\exp(-n^2/2\sigma_n^2)$.
Then 
\begin{eqnarray}
\langle \kappa^3 \rangle& = &\int_{-\infty}^{\infty}
\int_{-\infty}^{\infty} \kappa^3 
p(\kappa_s) p_n(\kappa-\kappa_s) d\kappa
d\kappa_s \nonumber \\ &=&
\int (\kappa_s+n)^3p(\kappa_s) p_n(n)d\kappa dn=
\langle \kappa_s^3 \rangle,
\label{momn}
\end{eqnarray}
since $\langle \kappa_s \rangle=\langle n\rangle=\langle n^3\rangle=0$. 
Note that by symmetry 
the latter relation still holds even if the ellipticity distribution is 
not Gaussian. Similarly we can correct $S_3$
so that it is unbiased (Hui \& Gazta\~ naga 1998). 

In contrast, for the other three statistics noise introduces bias
and $S_3$ is generally suppressed relative to the case of no noise.
This is because in the presence of noise
some $\kappa$ switch sign, which reduces the signal from these
statistics which are defined relative to the 0 mean. 
Unlike in the case of $\langle \kappa^3 \rangle $,
one has to correct for noise,
which requires knowledge of the full pdf to perform the integrals
such as in equation \ref{momn}. In practice this is best performed with 
N-body simulations.  If one corrects for noise bias one can see from 
figure \ref{fig7} that the optimal angular scale for the detection 
is around $1'-5'$. The detection 
is  at a several $\sigma$ level for the first method and a similar 
significance level is achieved if one combines the
second and third methods 
(which use two disjoint parts of the data set),
but at the expense of a more complicated noise corrections.
The fourth statistic fails completely in the presence of noise, as it 
gives the same value on small scales for all the models. 

To address the question of required survey size, we repeated
the above analysis for $2^{\circ} \times 2^{\circ}$ and 
$1^{\circ} \times 1^{\circ}$ fields. We find that 
$1^{\circ} \times 1^{\circ}$ field
is too small for the signal to be detected at more than a 1-sigma level.
The errors are significantly reduced for a $2^{\circ} \times 2^{\circ}$ 
field, which seems to be
the minimum required for a positive detection, at least 
for the survey depth and density we assumed here. This size is quite 
realistically achievable in the near future,
as there are several multi-square degree 
surveys under development which will reach the size 
and depth needed to measure this signal.

To summarize, both the third moment and the combination of the second 
moments over positive and negative values show clear
signature of a non-Gaussian signal even in the presence of noise.
The third moment $\langle \kappa^3 \rangle$ is the most sensitive 
and is unbiased in the presence of noise bias. On the 
other hand, it is somewhat more sensitive to numerical 
calibration errors than the 
two quadratic weighting methods. It is recommended that all the methods are
applied to the data and the results tested for consistency.
Moments of order higher than 3 are to be avoided because of sensitivity to 
the rare events in the tails and to uncertainties in the
numerical simulations. The lowest order method, based on the difference 
between negative and positive values, fails in the presence of noise.

\subsection{Comparison of results from PM and P$^3$M simulations}

\begin{figure}[p!]
\vspace*{18 cm}
\caption{Comparison between the pdf from P$^3$M (thin curves) and PM (thick 
curves) simulations for the same smoothing angles as in the previous 
two figures. The 
model shown is $\Omega_{\rm m}=0.3$, $\Omega_{\lambda}=0$, $\Gamma=0.21$ and 
$\sigma_8=0.85$.}
\includegraphics{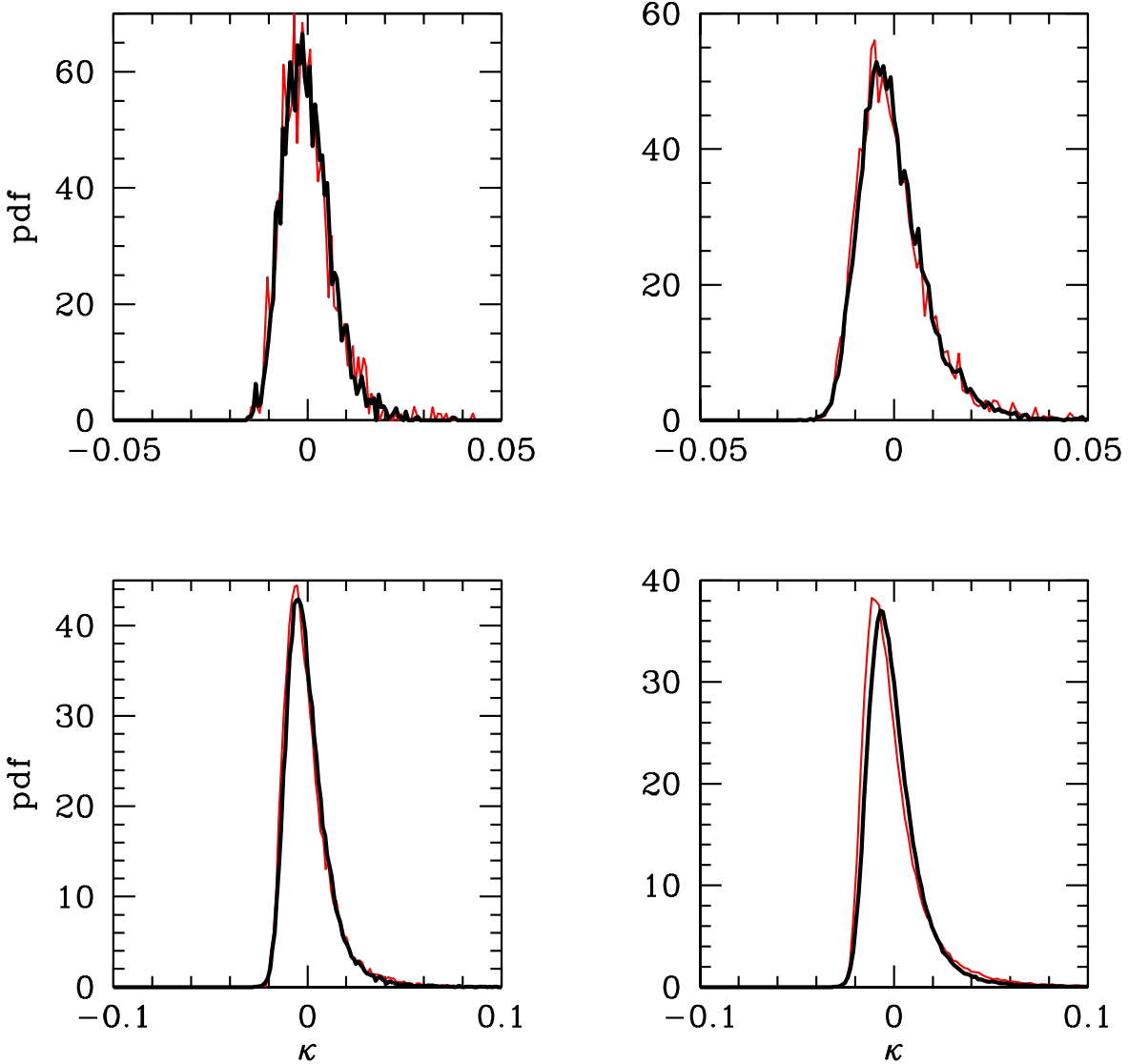}
\label{fig4}
\end{figure}

All the results shown above have used P$^3$M simulations. 
It is instructive to compare them to faster PM simulations,
which are expected to agree with P$^3$M simulations 
except in very dense regions. This is more important on small scales 
and/or higher moments. 
Figure \ref{fig4} shows the 
comparison between the pdf for PM and P$^3$M simulations for 
the $\Omega_{\rm m}=0.3$ model. 
While the P$^3$M simulation has $256^3$ particles and is one
of the highest resolution simulations existing at present, the PM
simulation used $128^3$ particles on a $256^3$ grid and only required
of order 10 CPU hours on a single processor workstation. 
Nevertheless, the agreement in the pdf is 
quite good for all the smoothing lengths and only on the smallest scales
does one see small differences. 
These results are reassuring and indicate that if one is using the pdf
away from the tails
the faster PM simulations suffice. However, the signatures in the tails
are more discriminatory, so it is important to test any statistic one
is using for its sensitivity to numerical inaccuracies. 

\begin{figure}[p!]
\vspace*{19 cm}
\caption{Comparison between P$^3$M (thin upper curves) and PM (thick 
lower curves) for the same statistics as in figure \ref{fig6}. The 
model shown is $\Omega_{\rm m}=0.3$, $\Omega_{\lambda}=0$, $\Gamma=0.21$ and 
$\sigma_8=0.85$.}
\includegraphics{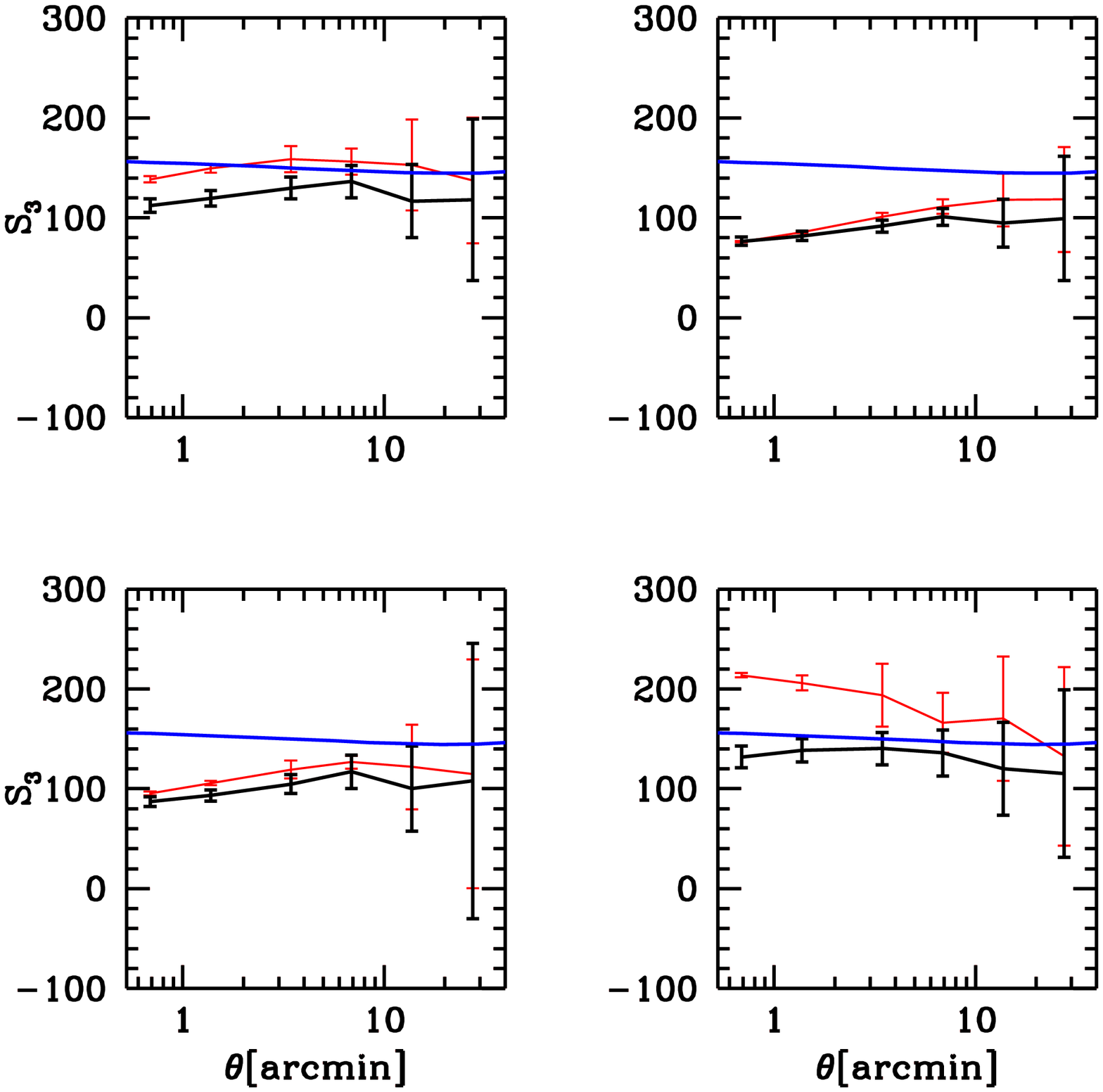}
\label{fig3}
\end{figure}

The comparison of results for $S_3$ between PM and  P$^3$M simulations 
is shown in figure \ref{fig3} for the $\Omega_{\rm m}=0.3$ model. 
On large scales both simulations
give results that are in agreement with each other and with perturbation
theory. On small scales differences are evident, 
more significantly for those measures that weight the positive tail of 
the pdf more strongly. This is especially true for
the third moment method (lower right), 
which is based on averaging $\kappa^3$ and so 
most strongly weights the high density regions. The difference between 
PM and P$^3$M in this case reaches a factor of 2 on small scales. 
The PM simulation is clearly inadequate for this statistic. In general
these results show that for higher moments the PM simulations have to
tested and calibrated with P$^3$M as the resolution range is difficult
to estimate a priori. One may worry that the P$^3$M
results may also be sensitive to numerical resolution on the 
smallest scales, so it is better to use statistics most sensitive to 
the tails on larger scales where the differences are smaller. 

\section{Likelihood analysis of the pdf}

The alternative to moments for probing non-Gaussianity is to use the
pdf of the reconstructed convergence. The pdf by definition describes the 
full distribution of $\kappa$ and so is not unduly sensitive to
the high density tails.
The simplest way to analyze the pdf is to use the Edgeworth
expansion (Juszkiewicz et al. 1995; Kim \& Strauss 1998), 
which expands the pdf around a Gaussian in a  
Hermite polynomial series. 
At lowest order one can fit for two free parameters,
the variance and skewness (equation \ref{edge}).
To obtain  the probability distribution for the measured $\kappa$ we convolve 
$p(\kappa)$ with the noise probability distribution
$p_n(n)=(2\pi \sigma_n^2)^{-1/2}\exp(-n^2/2\sigma_n^2)$. The 
convolution integral can be analytically calculated, giving the 
pdf for the measured $\kappa$
\begin{eqnarray}
&&p(\kappa| S_3,\sigma_{\kappa})=\int_{-\infty}^{\infty} 
p(\kappa_s) p_n(\kappa-\kappa_s) d\kappa_s =\nonumber \\
&&
{ 1 \over \sqrt{2 \pi (\sigma_{\kappa}^2+\sigma_n^2)}}
\exp\left[{-\kappa^2 \over 2 (\sigma_{\kappa}^2+\sigma_n^2)}\right] 
\nonumber \\ &&\left[
1+{1 \over 6} S_3\sigma_{\kappa}\left(H_3[\kappa/\sigma_{\kappa}(1+\alpha^2)]
+3 {\kappa \over \sigma_{\kappa}(\alpha+\alpha^{-1})^2} \right) \right],
\end{eqnarray}
where $\alpha=\sigma_n/\sigma_{\kappa}$. 
In the limit $\alpha \rightarrow 0$ noise is negligible and the equation
above reduces to the Edgeworth expansion as in equation \ref{edge}. In the 
limit $\alpha \rightarrow \infty$ it reduces to a Gaussian 
with $\sigma_n$ rms, in which case the non-Gaussian signature is
swamped out by the noise. 

To compute the likelihood function on the data 
we divide the observed region into 
nonoverlapping square cells and use the average value of $\kappa$
in each cell as the input.
The full likelihood function can be written as 
a product of likelihood functions for individual measurements $\kappa_i$,
\begin{equation}
L= \Pi_i \ p(\kappa_i| S_3,\sigma_{\kappa}).
\label{lik1}
\end{equation}
Individual $\kappa_i$ are not independent (because of 
correlations in the signal part of convergence; note that noise is
uncorrelated for reconstructed $\kappa$ provided that
one uses non-overlapping cells for analysis and the number of galaxies
in each cell is the same. The later is not exactly true because the galaxies
are Poisson distributed, so there are some residual noise correlations left, which
decrease as the cell size is increased), so 
this is not the true multivariate 
likelihood function for the data even if $p(\kappa_i| S_3,\sigma_{\kappa})$
were the correct pdf. 
Nevertheless we can continue to use it, as long as we 
determine the error estimates and possible biases with Monte Carlo 
realizations of N-body simulations. Given that this is not the
true likelihood function we have no reason to expect it to be an optimal 
estimator, even in the limit of large number of independent
measurements. One should really view it as another statistic whose 
performance should be judged empirically by the scatter in the Monte 
Carlo realizations. Since it uses information 
on the full pdf and not just its tails, one hopes
that it is a more powerful estimator than the moments. 

\begin{figure}[p!]
\vspace*{19 cm}
\caption{$S_3$ using ML (top) and $\chi^2$ (bottom) for the same
models as in figure \ref{fig6}. Noise has been added to the data.
}
\includegraphics{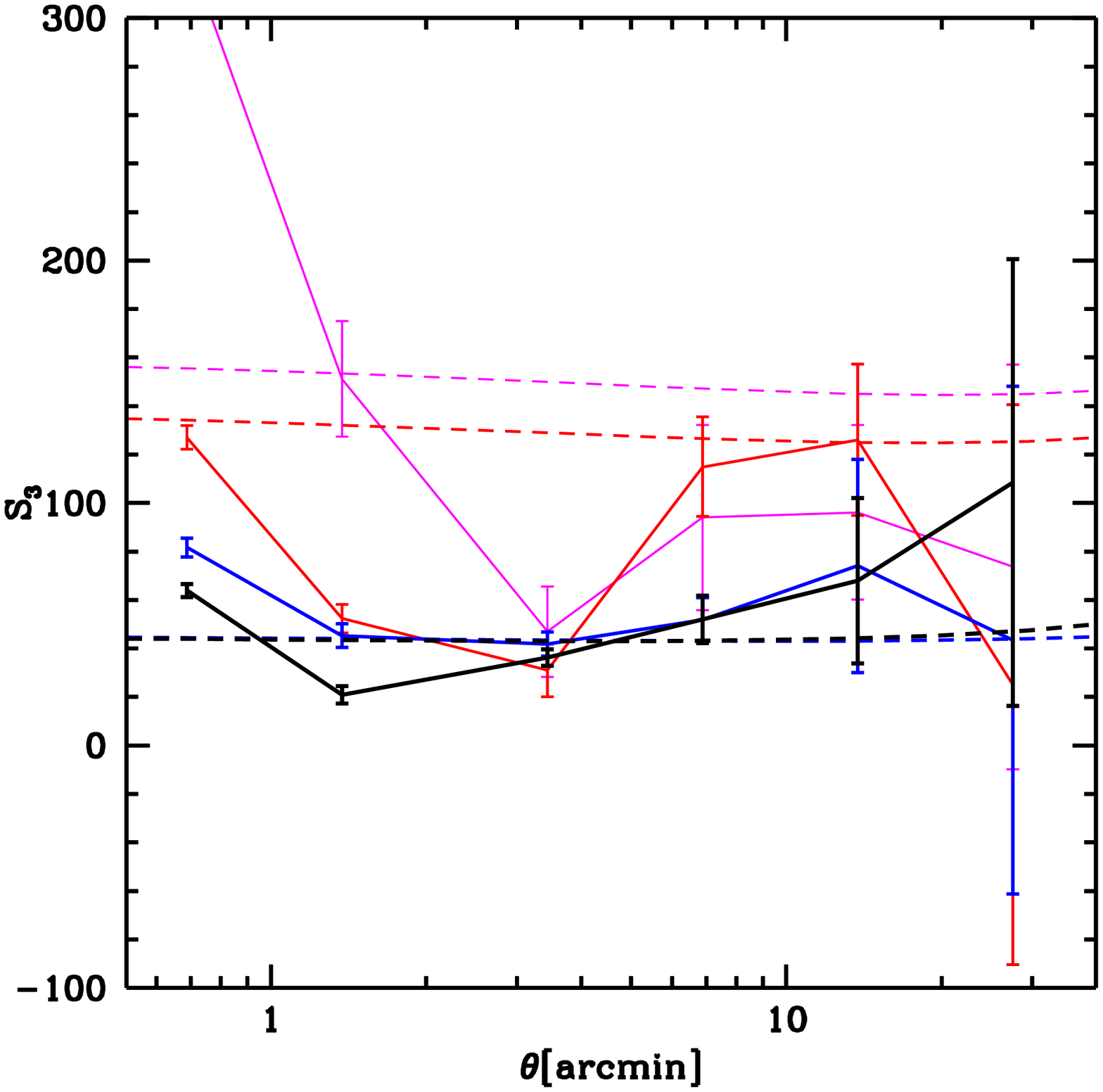}
\includegraphics{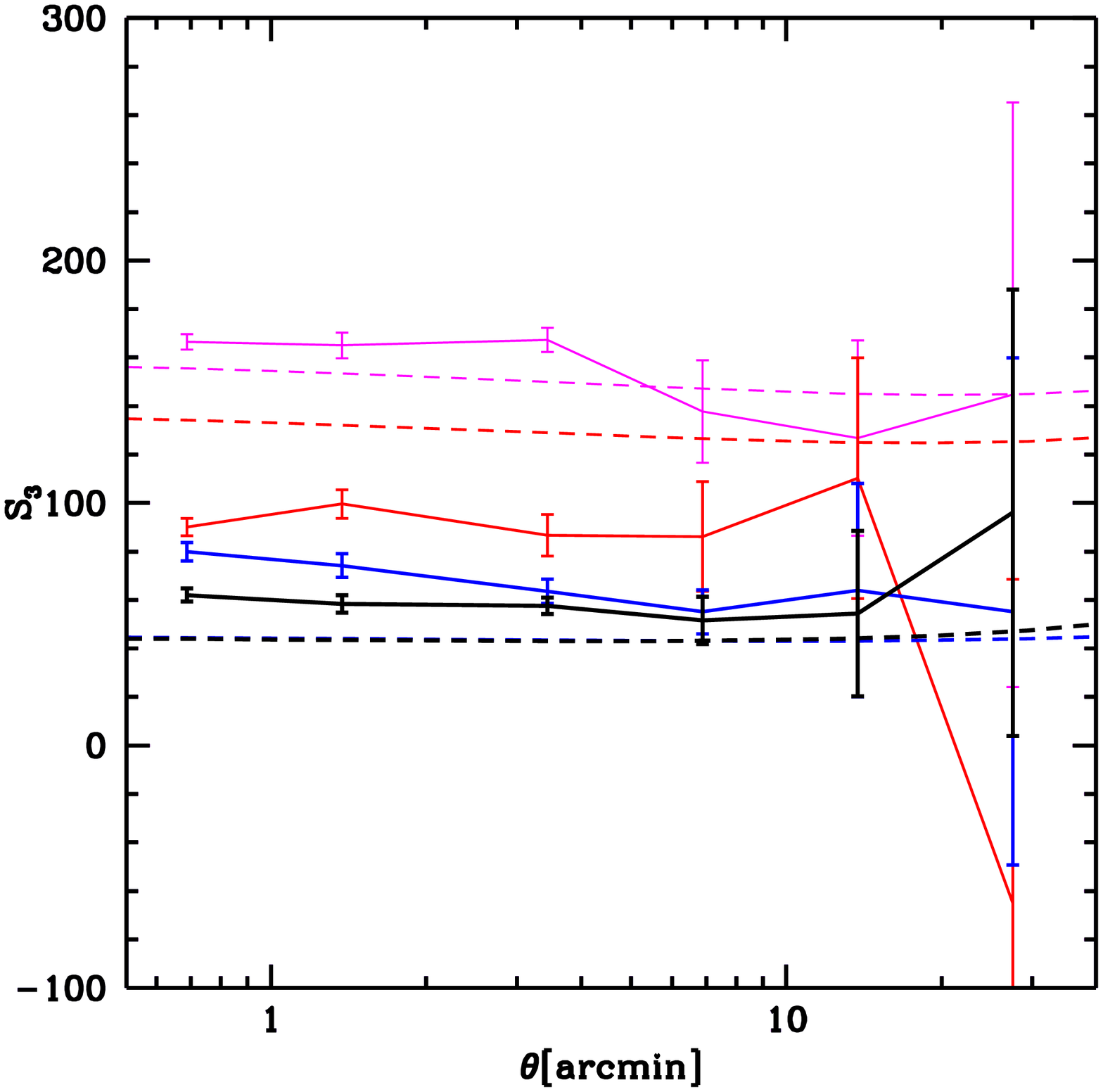}
\label{fig13}
\end{figure}

Figure \ref{fig13} (upper panel) shows the results of the ML method 
on the Edgeworth expansion 
for $S_3$ for various models and smoothing scales. The method works
well for the flat models, which have a lower $S_3$. It agrees with 
perturbation theory and has very small error bars even on small 
angular scales where noise is dominant. 
Visually, the agreement between the Edgeworth expansion 
and true noise convolved pdf is also quite good.
However, in low $\Omega_{\rm m}$ models 
there is a severe bias on intermediate scales, where the information
content is the highest from the moments analysis. 
The problem is that the Edgeworth expansion of the pdf does not 
satisfy the positivity condition for a probability distribution. 
It shows nonphysical oscillations because of the oscillatory 
nature of Hermite polynomials. For some values of $\kappa$
the pdf is not positive for large $S_3$, because the 
oscillations drive it below 0. This happens only over small regions, 
but if a measurement falls into one of these regions then $\ln(p)$
around it can have very large negative values and the ML method
will drive the estimate for $S_3$ and $\sigma$ away from this region 
into the region where the pdf is positive definite, even if visually 
the overall fit is much worse. Noise makes
the distribution more positive by smoothing out the nonphysical oscillations
(Kim \& Strauss 1998), but  in the present case, with realistic levels
of noise due to the finite number of galaxies, weak lensing 
data appear to be ``too good'' for a blind application of ML on the
Edgeworth expansion. This situation is somewhat paradoxical, because 
when the data are too good the method fails and one is forced to 
resort to  less accurate methods such as moments. 
The two ways to remedy this problem are: 
change the pdf to make it positive definite, or change the analysis 
method to allow negative pdf. We discuss both of these next.

\subsection{Likelihood analysis by fitting to simulations}

The failure of the ML method based on the Edgeworth expansion 
is caused by an unphysical expansion of the pdf. 
This suggests that it is better
to use directly the pdf from N-body simulations, which
provide more physical templates against which one can fit the observed 
data. By spanning a large parameter space of cosmological models one can 
search for the best model using the ML method. This seems an ambitious 
program, since it requires obtaining pdf's from N-body 
simulations of a dense grid of cosmological models. In practice it
is not so formidable, because a given simulation can be used for
many different models and because  one can use PM 
simulations, which are fast to compute and are 
sufficiently accurate away from the tails. 

To obtain a set of templates we first ran a number of PM simulations for 
different cosmological models. For each simulation we chose the shape
of the power spectrum (determined by $\Gamma=\Omega_{\rm m}h$ parameter in 
this case), final $\Omega_{\rm m}$ and final amplitude $\sigma_8$. 
A single simulation can 
be used not just for the chosen model, but also for models with smaller
$\sigma_8$. For example, in an $\Omega_{\rm m}=1$ model using
an output at redshift $z$ as the output today corresponds to a model with 
$\sigma_8^{\rm new}=\sigma_8^{\rm old}/(1+z)$. The shape of the 
power spectrum remains unchanged and for this 
case $\Omega_{\rm m}=1$ for all the models. 
For the case with curvature or cosmological constant earlier
time output corresponds to a higher $\Omega_{\rm m}$. Because the
shape parameter $\Gamma=\Omega_{\rm m}h$ is unchanged $h$ must then 
be decreased relative to the original
model. Thus a one-dimensional family of models can be 
spanned by the simulation. 

Another transformation that varies the model is a
rescaling of the box size. For example, using a 100$h^{-1}$Mpc simulation
as a 200$h^{-1}$Mpc simulation changes the shape of the power spectrum
so that $\Gamma$ is reduced by a factor of 2. At the same
time, for a given output $\sigma_8$ is increased, so one has to use an earlier
output to obtain the same normalization. It is therefore possible to use
a single simulation to describe a two-parameter family of models. Note
that the second transformation changes the angular resolution and size
of the simulation, so one has to be careful when comparing the results at
a given angular scale.
Since each PM simulation requires of order 10 CPU
hours on a workstation, 
one can span a dense grid of models in a reasonable amount of time.
Here we will present results as a two parameter family ($\Omega_{\rm m}$ and 
$\sigma_8$), although the third parameter $\Gamma$ can easily be added
if desired.

We again use the ML method to determine the best model using
\begin{equation}
L= \Pi_i\ p[\kappa_i| \Omega_{\rm m},P_{\delta}(k)],
\label{lik}
\end{equation}
where $p[\kappa_i| \Omega_{\rm m},P_{\delta}(k)]$ is the pdf for the model 
determined with $\Omega_{\rm m}$ and $P_{\delta}(k)$. These pdf's are 
obtained from the simulations, by binning them into small bins and 
counting the number of events in them. 
If there are $M_i$ simulated values of $\kappa$ 
contributing to the i-th bin then the pdf for that bin is 
given by $p_i=M_i/(\sum_i M_i)$. To obtain the log
likelihood for a given realization we count the number of $\kappa$ in 
the i-th 
bin, $N_i$, and compute $\sum_i N_i \ln(p_i)$. This expression shows
how the weight is distributed in the ML method. Around the peak of the
pdf there are many events and so $N_i$ is large, but $\ln(p_i)$ itself has
a small negative value. 
For the tails the number of events is small, but each event 
has a large weight, because $\ln(p_i)$ can have very large negative value.
This way both the peak and the tail of pdf are significant for the 
ML estimate. 

The main drawback with this approach is that 
one has to use very large simulations to properly sample the tails
of the pdf, so that the bins are appropriately populated. If this is 
not satisfied then even 
a single event in the real data falling into a bin that was underpopulated
completely dominates the rest of 
the data. Using ML we are therefore faced with the same problem we had with 
high order moments, which is that sensitivity to the tails implies 
large uncertanties because of numerical inaccuracies in simulations.
Clearly, using information from the rare events enhances the 
statistical significance of the test, but also increases the 
``systematic uncertanties'' that arise from the uncertainties in 
the N-body simulations. 

\subsection{$\chi^2$ fitting for the pdf}

An alternative to the ML method is a version of the $\chi^2$ statistic. 
$\chi^2$ fitting can be used both on the pdf from the 
simulations as a template or on the Edgeworth expansion. 
For example, we may compute the difference between 
measured and expected number of values in the i-th bin for a given model 
and add up the squares of these differences. We can then again minimize
this statistic as a function of the model and explore the errors on the 
parameters by Monte Carlo simulations. In addition, depending on how one
weights the differences one can make the central regions of the pdf more or 
less dominant compared to the tails. With the 
weighting according to the expected number of events, the tails become
more important and the method becomes similar to the ML method in how
it weights the data. No weighting makes the central region more important. 
Some intermediate weighting is therefore optimal 
if one wants to maximize both statistical significance and robustness to the
N-body results. 

We tried several simple weighting schemes and found that even with uniform 
weighting 
applied to the Edgeworth expansion, the $\chi^2$ statistic gave smaller error 
bars than moments (figure \ref{fig13}, lower panel). 
Moreover, $S_3$ was relatively 
constant as a function of angle and the difference between the
open $\Omega_{\rm m}=0.3$
and either of the flat models is enormous at $3'$. Both
flat models with different power spectra gave very similar predictions. 
Interestingly, the cosmological constant 
$\Omega_{\rm m}=0.3$ model is much closer to the flat model than 
to the open one, so it appears that this test is more sensitive 
to $\Omega_{\rm m}$ for open than 
for cosmological constant dominated models. Comparison between PM and 
P$^3$M simulations gave somewhat discrepant results, similar to those
in figure \ref{fig3},  
so it appears that this method is also sensitive to simulation resolution. 
Clearly, high resolution N-body simulations are essential if one is to 
extract the value of $\Omega_{\rm m}$ with high accuracy.

Finally, we also used the $\chi^2$ statistic on the pdf templates from
simulations. While this statistic is more accurate than all the statistics
previously discussed, the increased accuracy does not appear worth the
complexity and effort of simulating models spanning a large range of
parameters. For example, it is difficult to distinguish between 
an $\Omega_{\rm m}=0.2$  and $\Omega_{\rm m}=0.3$
model, in the sense that using the $\chi^2$ for one model was often 
better on the other template (with appropriate change in the normalization). 
The sensitivity therefore appears not to be significantly increased if one uses
the true pdf instead of the Edgeworth expansion. Given 
the increase in complexity of this method its usefulness may be
limited, at least until the data become of sufficient quality to
necessitate more accurate pdf's than provided by the simple first order 
Edgeworth expansion.

\section{Discussion}

In this paper we have used ray tracing simulations to study 
the following aspects of weak lensing by large-scale structure: 
the numerical requirements for simulating weak lensing on angular
scales of observational interest, $1'\lsim\theta\lsim 2^\circ$; 
the validity of various weak lensing approximations and of 
analytical predictions; computation of low-order statistics 
for different models and estimates of the accuracy with which they 
can be measured from given survey sizes; estimate of $\Omega_{\rm m}$ using
measures of non-Gaussianity from simulated noisy data. Our conclusions are 
as follows. 

\noindent{\bf Numerical tests and dynamic range from simulations.}
We have used FFT's to compute the shear matrix and photon deflections
required for propagating light rays in our simulations. The method
has been demonstrated to be efficient and accurate provided care is
taken with the choice of numerical parameters. We give analytical estimates
of the resolution limit implied by (a) the mass resolution and
force softening of the N-body simulation, and (b) grid effects in the
ray tracing. These estimates were shown to be reliable and can be
used to choose numerical parameters such as the grid size. 

Lens planes over a broad range of redshifts were found to contribute
to the lensing effect. For a given angular size, lens planes at 
lower-redshift have smaller spatial extent, of order 10 $h^{-1}$ Mpc for 
degree size fields. Since the lensing efficiency is a broad function of
redshift, non-negligible contributions come from
small regions that may contain a single over-
or under-density with only a small fraction of the particles in the 
full N-body simulation. This is a complication in achieving 
sufficient accuracy from ray tracing simulations. On small scales the white
noise contribution to the power spectrum from the finite number of 
particles is difficult to estimate as it depends on the clustering 
evolution, and therefore on the cosmological model. Given these 
limitations, to achieve a dynamic range $\gsim 100$ in angular scale, 
it is necessary to use high-resolution
N-body simulations with $\gsim 10^7$ particles, and FFT/ray tracing
grids of size $\gsim 1000^2$ at each lens plane. 

\noindent{\bf Nonlinear effects and comparison with analytical 
approximations.}
The power spectra measured from the ray tracing simulations 
(see figure \ref{figpower}) show that for fields a few degrees on 
a side the power spectrum of the shear lies entirely in the
nonlinear regime. Part of this nonlinearity arises because of the strongly
clustered regions at low redshift that affect statistical measures even
on degree scales. 
On scales of observational interest linear theory or
perturbation theory do not provide accurate predictions for 2nd and 3rd 
moments. 
The skewness measured from a reasonable number of degree sized 
fields is significantly enhanced over the perturbation theory
prediction. Only for the second moment of the shear and convergence
can non-perturbative analytical calculations be made 
(Jain \& Seljak 1997) and are found to agree well with the simulations. 
It is therefore necessary to use simulations in making theoretical 
predictions and obtaining error estimates for most weak lensing statistics. 

The weak lensing approximation is found to be valid to
good accuracy. The nonlinear departures discussed above arise from
gravitational clustering of the matter, not from any breakdown
of weak lensing approximations. There are occasional strong deflections
that contribute to the tails of the probability distribution of shear
and convergence. But lensing statistics smoothed on scales
larger than $1'$ do not receive significant contributions from these
rare events. An example of a test of the weak lensing approximation
is provided by figure \ref{figantisym}. The rotational component of
the distortion tensor, which arises when significant multiple 
deflections occur along the same line of sight, is shown to be very small. 
It is therefore possible to describe the shear and convergence
fields as arising from a projected (2-dimensional) gravitational potential. 

\noindent{\bf Sample size for measuring the power spectrum.}
The power spectrum of the shear directly probes the dark matter power 
spectrum at intermediate redshifts. Extensive 
tests have been used to estimate the dynamic range with which we
can measure the power spectrum and the accuracy on different scales.
For large fields and on large scales, sample variance (finite number of modes) 
dominated fluctuations
in the measured spectrum between different fields and can be reliably 
estimated using Gaussian statistics. On small scales this variance 
increases relative to the Gaussian value because nonlinear evolution generates 
kurtosis. In 
addition, power spectrum estimates from neighboring bins become strongly
correlated. For a quantitative analysis of the variance in the power spectrum 
it is essential to use numerical simulations. 
For small fields (below $1^{\circ}$ on a side) mode-coupling
becomes the limiting factor in the variance of power spectrum estimates rather
than number of modes. We found that for fields smaller than
1$^\circ$, it is possible to correct the power spectrum for fluctuations
in the mean convergence within every field. With the correction 
it is possible to reduce the error on the power by up to a factor of 2,
meaning that the errors are dominated by mode coupling on the scale
of the sample rather than number of modes within the sample. 
Measurements of about 10 fields a degree on a side can provide
power spectra accurate to a few tens of percent on $10^3\lsim l\lsim 10^4$.
This procedure depends on the knowledge of mean convergence in the field 
and its translation into mean density, which makes it an alternative, 
albeit less sensitive, test 
of $\Omega_m$ to the ones described in sections 6 and 7. 

\noindent{\bf Non-Gaussianity and $\Omega_{\rm m}$.}
We addressed the issue of how to extract information 
on the cosmological matter density $\Omega_{\rm m}$ from non-Gaussian 
signatures in weak lensing data. In the absence
of noise, the pdf of the convergence shows a number of characteristic
signatures of nonlinear evolution. In particular, high positive values of 
the convergence are characteristic of collapsed regions, while low negative
values show the presence of voids. The high density features depend both on 
the density parameter and somewhat on the shape of the power 
spectrum, though the dependence on the latter is reduced by smoothing the 
data on arcminute scales. The difference between the mean and the minimum 
convergence is in principle proportional to $\Omega_{\rm m}$. 
In our simulations the difference does grow as a function of 
$\Omega_{\rm m}$, but the effect is reduced somewhat because in high 
$\Omega_{\rm m}$ models structure is more linear at 
early times, so that there are virtually no empty beams out to $z=1$.

In the presence of noise the nonlinear signatures are masked and require 
careful calibration with N-body simulations to extract them. Based on an
analysis of several statistics we find that the most promising are: 
$\chi^2$ analysis on the Edgeworth expansion of the pdf,
the third moment, and  the second moment separately over positive and negative 
values, in decreasing order of statistical significance. 
The advantage of these measures is that they are simple, 
agree with perturbation theory on large 
scales and can be reasonably well calibrated using N-body simulations
on small scales. Methods based on a maximum likelihood  
fit to the Edgeworth expansion of the pdf 
improve the accuracy of the determination, but can give biased results. 
Methods based on maximum likelihood 
or $\chi^2$ fits to the pdf from N-body simulations are the most
accurate for a given model, but to span a range of models would 
require a large number of simulations for a relatively small gain in accuracy.

Noise due to the intrinsic ellipticities of galaxy images masks the
signal on very small scales. For the second and third moment the 
optimal regime is around a few arcminutes, while $\chi^2$ on the Edgeworth
expansion can give reliable results to significantly smaller scales.  
With a several degree sized survey of reasonable depth one can 
determine $\Omega_{\rm m}$ to within about 0.1-0.2 for
open models, and somewhat worse for cosmological constant dominated
models, based on the statistical errors. The
accuracy is also limited by systematic effects, 
such as the shape and amplitude of the power spectrum. These effects 
can be modeled with a power spectrum analysis of the same data.
Another source of systematics is calibration with 
N-body simulations. Based on comparisons between PM and P$^3$M simulations 
we find that the calibration with N-body simulations is relatively secure 
on a few arcminute scales, while on smaller scales the differences
for some of the statistics are quite significant.
This is particularly true for the third moment, which is very sensitive
to the tails of the distribution. For this reason we did not explore
moments higher than the third, such as kurtosis,
which would be even more sensitive to the resolution of the simulation.
Thus well calibrated, high resolution simulations are essential to probe 
the high density tails of the pdf. 
Finally, detecting non-Gaussianity in the weak lensing signal
also provides a direct detection of nonlinear 
evolution of the dark matter. If the results from such an analysis are found 
to agree with other tests it would provide strong support for the 
gravitational instability paradigm.

\section*{Acknowledgments}
We are grateful to Anthony Banday, Matthias Bartelmann, Micol Bolzonella, 
Lauro Moscardini, Ue-Li Pen, Peter Schneider,
Alex Szalay, Ludovic van Waerbeke, Joachim Wambsganss and Matias 
Zaldarriaga for useful discussions. We thank an anonymous referee for
helpful suggestions. 
The high resolution simulations in this paper 
were carried out using codes made available by the Virgo consortium. 
We thank Joerg Colberg for help in accessing this data
and Ed Bertschinger for making available his PM N-body code.
BJ acknowledges support from NASA through the LTSA grant NAG 5-3503.

\end{document}